\pdfoutput=1
\documentclass[showpacs,singlecolumn,superscriptaddress]{revtex4-1}

\usepackage{amsmath,amssymb}
\usepackage{graphicx}
\usepackage{dcolumn}
\usepackage{bm}
\usepackage{color}
\usepackage[normalem]{ulem}

\begin{document}
\title{The role of Hurst exponent on cold field electron emission from conducting materials: from electric field distribution to Fowler-Nordheim plots}

\date{\today}

\author{T. A. de Assis}
\address{Instituto de F\'{\i}sica, Universidade Federal da Bahia,
   Campus Universit\'{a}rio da Federa\c c\~ao,
   Rua Bar\~{a}o de Jeremoabo s/n,
40170-115, Salvador, BA, Brazil}
\address{Instituto de F\'\i sica, Universidade Federal Fluminense, Avenida Litor\^anea s/n,
24210-340 Niter\'oi RJ, Brazil}
\email{thiagoaa@ufba.br}

\begin{abstract}

This work considers the effects of the Hurst exponent ($H$) on the local electric field distribution and the slope of the Fowler-Nordheim (FN) plot when considering the cold field electron emission properties of rough Large-Area Conducting Field Emitter Surfaces (LACFESs). A LACFES is represented by a self-affine Weierstrass-Mandelbrot function in a given spatial direction. For $0.1 \leqslant H < 0.5$, the local electric field distribution exhibits two clear exponential regimes. Moreover, a scaling between the macroscopic current density ($J_M$) and the characteristic kernel current density ($J_{kC}$), $J_{M} \sim  \left[J_{kC}\right]^{\beta_{H}}$, with an H-dependent exponent $\beta_{H} > 1$, has been found. This feature, which is less pronounced (but not absent) in the range where more smooth surfaces have been found ($0.5 \leqslant H \leqslant 0.9$), is a consequence of the dependency between the area efficiency of emission of a LACFES and the macroscopic electric field, which is often neglected in the interpretation of cold field electron emission experiments. Considering the recent developments in orthodox field emission theory, we show that the exponent $\beta_{H}$ must be considered when calculating the slope characterization parameter (SCP) and thus provides a relevant method of more precisely extracting the characteristic field enhancement factor from the slope of the FN plot.

\end{abstract}

\maketitle

{\large \bf Introduction }
\\
\\
The theory of cold field electron emission (CFE) is relevant for understanding the important technological aspects in the development of large-area electron emitter devices with apex radii of no less than a dozen nanometers \cite{Mayer}. This theory can be conveniently formulated using a free-electron model for the emitter and a Schottky-Nordheim barrier model for the tunneling barrier \cite{Forbes,Forbes2,Forbes3}.
Large-area field emitters provide an effective scenario for complying with the requirement that electron field emission in metals occurs when a large local electric field  ($\sim$ a few V/nm) induces electron tunneling through a potential barrier out of the corresponding surface into vacuum. Therefore, practical research on CFE involves the development of techniques to reduce the barrier through which the electrons must tunnel \cite{Zhuo, Cabrera}. For example, an interesting phenomenon of electron emission at low macroscopic electric fields has been reported for many materials, such as amorphous diamond-like films \cite{Geis}. Diamond cold cathodes formed using chemical vapor deposition (CVD) for use in field emission displays have attracted considerable attention because of their low work function \cite{Gu,Chartterjee}.  Moreover, the local field distributions over emitting surfaces have been explored, for example, using carbon-nanofiber (CN) films constructed of highly uniform nanofiber arrays; in that case, experimental evidence of Gaussian behavior of the distribution of the spatially resolved field enhancement factor was obtained via scanning anode field emission microscopy (SAFEM) \cite{Cole}. In addition, experimental results obtained using the same technique (measuring the constant macroscopic field emission current map) have revealed exponential decay in thin films of non-oriented carbon nanotubes (CNTs), which exhibit an irregular morphology on the corresponding surfaces \cite{Groning}.
Indeed, it is becoming increasingly clear that modified Fowler-Nordheim-type (FN-type) equations must be developed to address CFE from rough thin-film emitters (for example, films of CNTs and related materials) \cite{Liang}. In such a case, upon the application of a voltage to the support tips, a high local electric field develops at the apex, inducing a field emission tunneling current \cite{Hamri}. In fact, no definitive treatment has yet emerged for CFE from conducting materials that accounts for roughness at small scales. In these circumstances, improving the Fowler-Nordheim-type equation to yield a satisfactory fit to the experimental results is an important requirement for the development of field emitter devices with more precise specifications.

Only a relatively small number of studies have considered the role of the irregular morphology of the surface of a thin film on the local electric field distribution \cite{Lazic, Zadin} and the corresponding field electron emission properties of Large-Area Conducting Field Emitter Surfaces (LACFESs) in a genuine three-dimensional problem \cite{deAssis1, deAssis2}. However, considering the morphologies of the conducting surfaces that are commonly experimentally investigated, a large number of problems exist in which such a consideration is relevant \cite{Jeong, Arancibia}. Irregular morphological features that remain on the surface after the manufacturing process may act as field emitting tips in the presence of an external electric field. Indeed, a more detailed investigation of the surface preparation prior to the application of a high electric field leads to a considerable decrease in the breakdown rate \cite{Andrade}.

Motived by the aforementioned studies, in this work, we present the first consideration of the role of the Hurst exponent $H$ of an irregular LACFES in determining the related local electric field distribution and field electron emission properties using orthodox field emission theory. A LACFES is represented by a self-affine Weierstrass-Mandelbrot (WM) function in a given spatial direction such that all surfaces are assumed (i) to exhibit no roughness along any other spatial direction and (ii) to have the same global roughness (which is a measure of the fluctuations at large scales). The latter condition allows us to systematically evaluate the role of $H$ (and the short-wavelength fluctuations in the morphology of the LACFES) in determining the non-linearities that may appear in the corresponding Fowler-Nordheim (FN) plots.

The results indicate that for any $H$, the local electric field distribution over a LACFES exhibits two exponential regimes, implying a non-linear (power-law) relation between the macroscopic current density and the characteristic current density, if orthodox field electron emission is assumed. We show that this scaling is a consequence of the dependence between the area efficiency of emission of the LACFES and the macroscopic electric field, which is more pronounced for $ 0.1 \leqslant H < 0.5$. This result allows for the introduction of a new slope correction parameter in an FN-type equation to allow for the more precise extraction of the characteristic field enhancement factor (FEF) from the slope of the FN plot.
\\
\\
{\large \bf Morphology of a LACFES}
\\
\\
Central to our approach is an expression that represents the height profile of the emitter along a line on the emitter surface (defined here as the $x$ coordinate). The other coordinate parallel to the emitter surface is denoted by $y$, and the coordinate normal to the surface is denoted by $z$. To model this profile, we use the Weierstrass-Mandelbrot (WM) function \cite{Berry}:

\begin{equation}
f_{WM}(x)\equiv c \sum\limits_{n=0}^{N} \xi^{-{Hn}} \sin(K_{0}\xi^{n}x + \phi_n),
\label{Eq1}
\end{equation}
where $c$ is a constant related to the global roughness amplitude, $0<H<1$ is the Hurst exponent parameter (also called the ``local roughness exponent" ($\alpha_l$) in experimental thin-film science), $K_{0}$ is the fundamental wavenumber, and $\phi_n$ is an arbitrary phase (in this work, we consider the value of $0 < \phi_n < 2 \pi$ to be randomly chosen). $\xi$ is a parameter ($\xi> 1$) which is a measure of the distance between the frequencies. This function can be used to model the emitter height profile as a superposition of a set of sine waves, with the wave amplitude decreasing as $\xi$ increases.

Larger values of the Hurst exponent imply a smoother surface because the ratio of the logarithm of the local roughness (Eq. (\ref{Eq1b}) below) to the logarithm of the scale length is larger (here, the scale length refers to the scale at which the statistical properties, such as height fluctuations, of an irregular surface are analyzed; i.e., the box size ``r" as a measure of the local roughness is defined in Eq. \ref{Eq1b}). For $0.5 <H< 1$, the behavior is called persistent, whereas for $0 <H < 0.5$, the behavior is called is antipersistent. Antipersistence indicates that the heights of various points on the surface are negatively correlated and that the correlations rapidly decay to zero. This scenario corresponds to a small ratio of the local roughness with respect to the scale. Finally, at $H=0.5$, the heights of the LACFES are statistically random. If we use the one-dimensional correlation function $\Omega(x)$, defined as

\begin{equation}
\Omega(x) = \frac{ \overline{-f_{WM}(-x) f_{WM}(x)}}{\overline{[f_{WM}(x)]^2}},
\label{Eq1aa}
\end{equation}
where the overbars represent spatial averages, it is trivial to show that $\Omega(x)$ does not depend on $x$. In fact, using Eqs. (\ref{Eq1}) and (\ref{Eq1aa}), $\Omega = \left(2^{2H-1} - 1\right)$. Thus, the signal of $\Omega$ defines the type of correlations that appear in the WM function, in such that $\Omega>0$, $\Omega<0$ and $\Omega=0$, for $H>0.5$, $H<0.5$ and $H=0.5$, respectively. The WM function is a combination of periodic functions that exhibits two major features: (i) it can be anisotropic, and (ii) it is self-affine with complexity at small scales. Thus, the consideration of such surfaces is motivated by the fact that real irregular surfaces are generally neither purely random nor purely periodic.

\begin{figure}
\includegraphics [width=10.5cm] {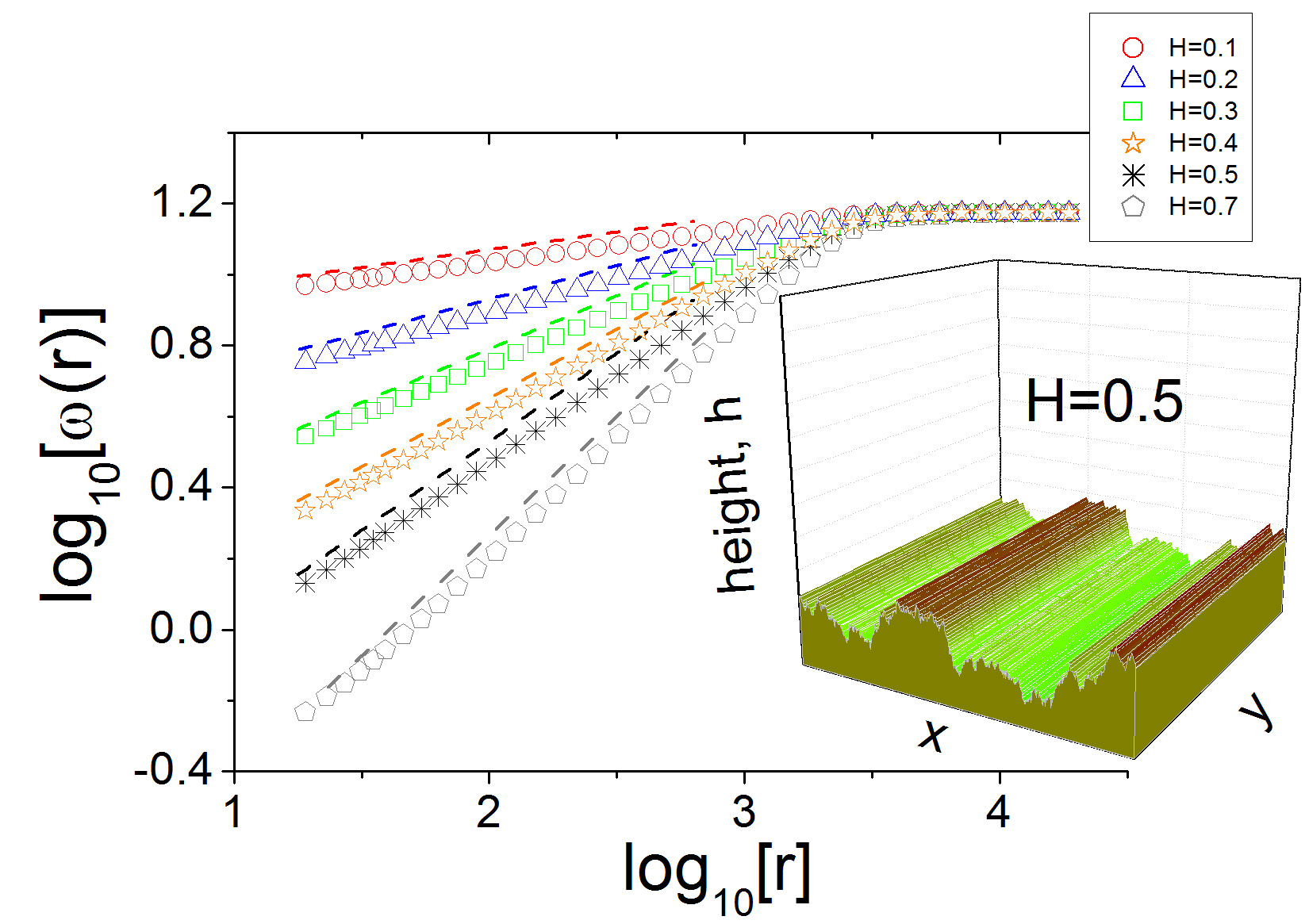}
\vspace{1.0cm}
\caption{Scale of the local roughness for the WM function with $H \in [0.1,0.7]$. The very good agreement between the slope ($\alpha_l$) of $\log_{10}[\omega(r)]$ as a function of $\log_{10}[r]$ and the parameter $H$ is remarkable. From top to bottom, the dashed lines correspond to slopes of $\alpha_l = 0.1, 0.2, 0.3, 0.4, 0.5$ and $0.7$. In the inset, a Large-Area Conducting Field Emitter Surface (LACFES) represented by a self-affine WM function (see Eq.(\ref{Eq1})) with the parameter $H=0.5$ and $\xi = 0.5$e is shown, where ``e" is Neper's number. It is evident that the conducting surface lies on a two-dimensional substrate with no roughness along the $y$ direction.} \label{Fig1}
\end{figure}
Now we use the notation $h \equiv f_{WM}(x)$ to represents the height of the profile in the $z$ direction with respect to the plane $h(x,y)=0$. The global roughness parameter ($W$) provides a measure of the height variation in the emitter profile, taken across the emitter as a whole. This parameter is defined as the root mean square of the front-surface height fluctuations and is given by

\begin{equation}
W \equiv \left[\overline{\left(h-\overline{h}\right)^{2}}\right]^{1/2},
\label{Eq1a}
\end{equation}
where the averaging is performed over a length $L_{x}$ that represents the lateral size of the emitter in the $x$ direction. To help determine whether the WM function provides a realistic and useful description of self-affine one-dimensional profiles acquired from real emitter surfaces, we can analyze how the local roughness $\omega(r)$ scales with the length $r$, which is a useful measurement for experimental purposes. This is also defined as a root-mean-square height fluctuation, but the spatial average is limited to a scale of size $r$, with $r < L_{x}$. More explicitly, the local roughness is given by

\begin{equation}
\omega(r) \equiv \langle [\overline{\left(h-\overline{h}\right)^{2}}]^{1/2} \rangle,
\label{Eq1b}
\end{equation}
where the angular brackets represent the configurational average obtained as the box (of length $r$) scans the entire irregular surface.

The function $f_{WM}(x)$ has no characteristic length (beyond the size of the system itself), in the sense that the level of detail of this function is self-similar under an affine scaling (in which the ``x" axis is stretched by a factor of $\xi$ and $f_{WM}(x)$ is modified by $\xi^{H}$, i.e., $f_{WM}(\xi x) = \xi^{H} f_{WM}(x)$). Therefore, its statistical properties are identical at different scale lengths. As indicated by the analysis presented in Fig. \ref{Fig1}, for small scale lengths, the local roughness scales as $\omega(r)\sim r^{\alpha_l}$, which is in contrast with large scale lengths, for which the local roughness coincides with the global $W$. The very good agreement between the slope $\alpha_l$ (from $\log_{10}[\omega(r)]$ as a function of $\log_{10}[r]$) and the parameter $H$ can be clearly observed. In the inset of Fig. \ref{Fig1}, the surface that mimics the LACFES in the case of $H=0.5$ is shown; this surface was computed using $N=200$ in the sum defined in Eq. (\ref{Eq1}). We stress that $x$ is considered to be an integer number that represents a distance measured in terms of some basic unit distance $u$, as are $y$ and $z$.

\begin{figure}
\includegraphics [width=12.5cm] {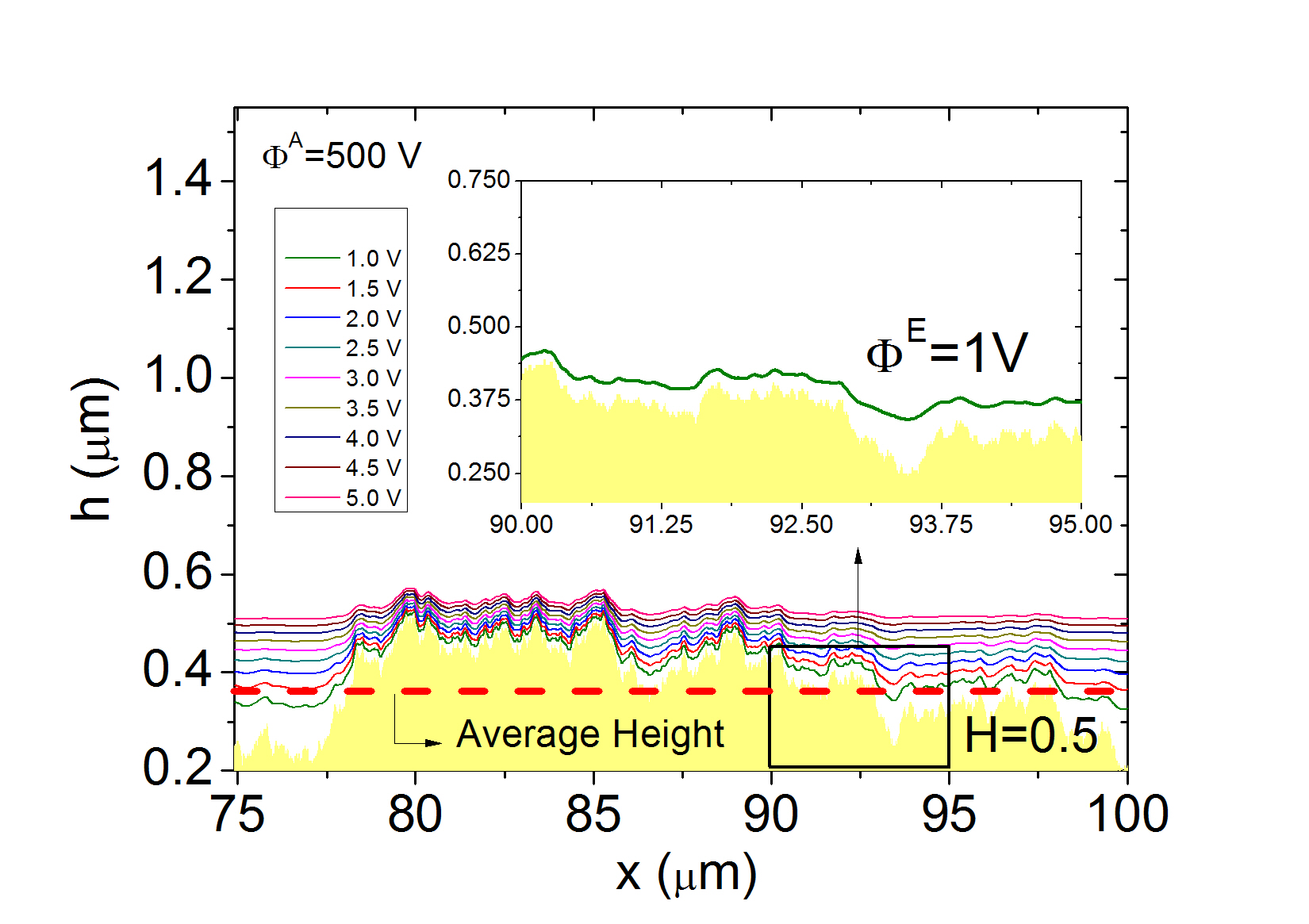}
\caption{
Profile of a LACFES with H=0.5 (shown in Fig.  \ref{Fig1}) and equipotential lines calculated numerically from the solution of the Laplace equation with appropriate Dirichlet conditions ($\Phi^{S} = 0$ V and $\Phi^{A} = 500$ V - see the text for more details). In the inset, a portion of the LACFES and the equipotential line defined by $\Phi^{E} = 1$ V, representing the approximation of the field emitter surface, are highlighted. The horizontal dashed (red) line indicates the average height of the roughness profile. The far-away conducting anode (not shown) is located at $h \approx 16 \mu$m from the average height of the LACFES (see the text for more details).} \label{Fig2}
\end{figure}
In fact, the typical Fowler-Nordheim-type theory of CFE was not developed at the atomic scale and thus does not apply to emitters that are sharp on that scale. To overcome this difficulty, we developed an algorithm to round off the sharpest projections and replace the real WM surface with an equipotential surface, $\Phi^{E}$ (very similar to the previous one), that is calculated from the numerical solution of the Laplace equation using a finite-difference scheme (Liebmann method - see Supplementary Information I). For practical applications, this approximation corresponds, for instance, to the condition in which good Spindt arrays \cite{Spindt} are produced by slowly increasing the voltage such that the sharpest tips ``burn off." Our numerical solution has been shown to yield an FEF for ideal protuberances that is in very good agreement with the results of previous analytical, finite-element and multigrid methods \cite{deAssis3,Djurabekova,deAssis4}.

In Fig. \ref{Fig2}, several equipotential lines are shown that consider electric potentials for the LACFES (for the case of $H=0.5$) and the far-away conducting anode (located at $h \approx 16 \mu$m from the average height of the LACFES) of $\Phi^{S} = 0$ V and $\Phi^{A} = 500$ V, respectively, corresponding to Dirichlet conditions. The equipotential surface, on which the local electric field distribution will be calculated, is shown in the inset of Fig. \ref{Fig2}. In this work, we use $u = 5$ nm, which results in a global roughness of $W \approx 75$ nm for all LACFESs. Moreover, this methodology ensures that for small scales (though larger than the atomic scale, e.g., $r = 20 nm$), the local roughness is negligible, and for large scales (i.e., $r > 500$ nm, although still considerably smaller than the lateral size of the system $L_{x}$), the local roughness scales with the same exponent $H$. The same procedure is adopted for any $H$ that allows for the use of CFE theory.
\\
\\
{\large \bf Orthodox CFE Theory - Local emission current density of a particular lateral location on a LACFES}
\\
\\
The recent developments in CFE theory can be viewed as a relevant approach for understanding the field emission properties of a rough LACFES.
The local emission current density (LECD), $J_L$, in the CFE regime can be written as a function of a convenient set of parameters, which includes the sum of all traveling states incident on the inside of the emitter surface considering all occupation and transmission probabilities. The general result takes the following form \cite{LBForbes}:

\begin{equation}
J_{L} = Z D,
\label{Eq1c}
\end{equation}
where $Z$ is an effective incident current density and $D$ is a transmission probability. In the case of high electric fields and low temperatures, the value of $D$ at a specific forward energy level (often that for a Fermi-level electron moving normal to the emitter surface) is chosen, and $Z$ is calculated as $J_{L}/D$. CFE is an LECD regime in which most electrons escape via deep tunneling (i.e., tunneling well below the top of the barrier) from states close to the emitter Fermi level, and the Landau and Lifschitz approximation can be applied \cite{LBForbes}. Then, using this approximation, Eq. (\ref{Eq1c}) can be written as follows:

\begin{equation}
J_{L} = Z_{F}P_{F} \exp{[-G_{F}]},
\label{Eq1d}
\end{equation}
where ``$F$" is used to indicate parameters related to a barrier of zero-field height ($\phi$). $G_{F}$ is the related barrier strength (also called the Gamow exponent or the JWKB (Jeffreys-Wentzel-Kramers-Brillouin) exponent) and is given by

\begin{equation}
G_{F} = 2 \varphi \int M^{1/2}(z) dz \equiv g_{e} \int M^{1/2}(z) dz.
\label{Eq1d1}
\end{equation}
In Eq. (\ref{Eq1d1}), $\varphi^{2} \equiv 2m_{e}/\hbar^{2}$, where $m_{e}$ is the electron mass and $\hbar$ is related to Planck's constant; $M(z)$ specifies the form of the barrier, which is the difference between the total electron potential energy, $U(z)$, and the total energy component, $E_{z}$, related to the motion of the electron in the z direction (perpendicular to the surface of the emitter in the case of a conducting emitter), $M(z)\equiv U(z) - E_{z}$. Moreover, the integral in Eq. (\ref{Eq1d1}) is performed over the range of $z$ for which $M(z)>0$.

Considering metal emitters with appropriate dimensions, as previously discussed, the LECD $J_{L}(\phi,F)$ is formally given by the following equations:

\begin{equation}
J_{L}(\phi ,F) = \lambda_{CFE} J_{k}(\phi ,F),
\label{Eq1e}
\end{equation}

\begin{equation}
J_{k}(\phi ,F) = a \phi^{-1} F^{2} \exp{\left[-\nu_{F}b\phi^{3/2}/F\right]},
\label{Eq1f}
\end{equation}
where $J_{k}(\phi,F)$ is the kernel current density, which can be evaluated exactly for any chosen barrier form, $\nu_{F}$, and choice of $\phi$ and $F$; $a$ (=$1.541434 \times 10^{-6}$ A eV $V^{-2}$) and $b$ (=$6.830890$ $eV^{-3/2}$ V $nm^{-1}$) are the first and second Fowler-Nordheim constants, respectively \cite{LBForbes}.

In this work, we consider the Schottky-Nordheim (SN) barrier used in the Murphy-Good theory \cite{Murphy} and, more recently, in the Forbes approximation (orthodox emission theory) \cite{Forbes}. The SN barrier corresponds to the lowering of the top of the barrier relative to that of an exactly triangular barrier such that $M(z) = \phi - eFz - e^2/16 \pi \epsilon_{0} z$ ($e$ is the positive elementary charge and $\epsilon_{0}$ is the electric constant). In this way, it is possible to define the scaled barrier parameter, $f$, which is given by

\begin{equation}
f = F/F_{R} = (4 \pi \epsilon_{0}/e^{3} \phi^{2}) F.
\label{Eq1d2}
\end{equation}
The reference field $F_{R}$ is the field that is required to lower the barrier height by an amount equal to the local work function $\phi$. For any $\phi$, it is possible to define a parameter $\eta(\phi)$ as follows:

\begin{equation}
\eta(\phi) = b \phi^{3/2}/F_{R} \approx 9.8362 (eV/\phi)^{1/2}.
\label{Eq1d3}
\end{equation}
If we compare Eqs. (\ref{Eq1d}) and (\ref{Eq1e}), the pre-factor $P_{F}$ used in the first equation is included in $\lambda_{CFE}$. One problem with attempting to obtain good predictions of the CFE current density is that exact values of $\lambda_{CFE}$ (which depends on the material) are not well known. The current best guess is that $\lambda_{CFE}$ lies in the range $0.005< \lambda_{CFE} <11$ \cite{LForbes}. We compute the local current density that results in $\lambda_{CFE}=1$ in Eq. (\ref{Eq1e}) such that $J_{L}(\phi,F) = J_{k}(\phi,F)$. In any case, the correction factor with the largest influence on $J_{L}$ is the barrier shape correction factor $\nu_{F}$, which will also be considered in this work. Finally, we use an approximation such that each point on the LACFES represents a particular lateral location on the emitter surface.
\\
\\
{\large \bf Results and discussion}
\\
\\
{\large \bf Local electric field distributions}
\\
\\
Before discussing the results for the LECD, we analyze the behavior of the local electric field intensity distribution over the LACFES, $\rho(F)$. In Fig. \ref{Fig3} (a), the behavior of $\rho(F)$ is shown for a LACFES with $H=0.1$ and for several values of the anode electric potential $\Phi^{A}$. The vertical black dashed line (gray dashed-dotted line) represents an electric field intensity of $2.5$ V/nm (4 V/nm - see Ref. \cite{ForbesResub}), which corresponds to a typical value for field emission in pure metals with a local work function of approximately $3.5$ eV ($4.5$ eV). The inset of this figure shows the behavior of the mean electric field $\langle F \rangle$ over the LACFES, which is defined by

\begin{equation}
\langle F \rangle \equiv \frac{1}{A_M} \sum\limits_{i,j} F_{i,j} \Delta x_{i} \Delta x_{j},
\label{Eq2}
\end{equation}
as a function of $\Phi^{A}$ for $H \in [0.1,0.9]$. Here, $A_M$ is the total ``substrate footprint" area, $F_{i,j}$ is the local electric field intensity at a particular location $(i,j)$ on the LACFES, and $\Delta x_{i} \Delta x_{j} = u^2$ corresponds to an area of unity, with $\Delta x_{i},\Delta x_{j} \ll L_{x}$. The results clearly suggest a linear dependence whose slope, as a non-linear function, depends on the $H$ of the LACFES. It is interesting to observe the approximate collapse of the $\langle F \rangle$ behavior for $0.5 \leqslant H \leqslant 0.9$. This allows us to more easily identify the primary differences between the electrical properties of a rough correlated (and random) LACFES ($0.5 \leqslant H \leqslant 0.9$) and a rough anti-correlated LACFES ($0.1\leqslant H<0.5$).

As a measure of emitter sharpness, we compute the distribution of the macroscopic field enhancement factors (FEFs), $\gamma \equiv F_{i,j}/F_{M}$, that are typically applied in measurements using scanning anode field emission microscopy. A high FEF factor leads to a low turn-on voltage and a high emission current, which are desirable for emitter applications. In these calculations, the field at the position of the average height of the LACFES ($\approx 0.37\mu$m for a LACFES with any $H$ - see Fig. \ref{Fig2} for $H=0.5$), in the absence of protrusions, was considered to be the macroscopic electric field $F_{M}$. Thus, the flat anode, which is placed at a height of $h^A$ (measured with respect to the substrate on which the film was grown, i.e., the plane $h(x,y)=0$), is located at a distance of $\langle d \rangle \equiv h^A - \overline{h}$ from the cathode. The macroscopic electric field, $F_{M}$, is given by

\begin{equation}
F_{M} = \frac{\Phi^A}{\langle d \rangle}.
\label{Eq2s}
\end{equation}

If $\Phi^A$ is taken to be $500$ V and $\langle d \rangle \approx 16 \mu$m (see Fig. \ref{Fig2}), then the macroscopic electric field is $F_{M} \approx 31$ V/$\mu$m. In Fig. \ref{Fig3}(b), the results for the distribution of $\gamma$, $\rho(\gamma)$, are shown for $H \in [0.1,0.9]$. This measurement can be regarded as a probability distribution because the number of field emitter locations per unit area on the LACFES can be written as $dN = \rho(\gamma) d\gamma$, where $dN$ is the number of locations with FEF $\gamma$ $\in$ [$\gamma$, $\gamma + d\gamma$] per unit area. For a given value of $H$, the corresponding distributions $\rho(\gamma)$ were found to be the same for any $\Phi^{A}$, indicating that these results are not dependent on $\Phi^{A}$. This is the result that is physically expected because in classical electrostatics (in the case of a diode), the value of the field enhancement factor depends only on the electrode geometry and not on the applied electrostatic potential difference. However, it is a useful test to confirm that this expected physical result is reproduced in our numerical simulations. An interesting alternative to this behavior is to observe a collapse of the local electric field intensity distributions $\rho(F)$, which is expected to occur if the variable $F$ is replaced by a scaled variable $\chi \equiv F/\langle F \rangle$. This phenomenon can be clearly observed in Fig. \ref{Fig3}(c), which presents the results for $H=0.1$ and for all $\Phi^{A}$ values shown in Fig. \ref{Fig3}(a). This collapse was also verified for all $H$ values considered in this work.

\begin{table}
   \centering
 \renewcommand{\arraystretch}{1.5}


 \begin{tabular}{|c|c|c|c|}

  \hline
 $H$ & $\delta_{H}^{I}$ $\times \log_{10} (e)$ & $\delta_{H}^{II}$ $\times \log_{10} (e)$ & $\gamma_{C}$ \\

 \hline

 0.1 & $0.343 \pm 0.001$ & $0.543 \pm 0.001$ &  $\approx 11.86$   \\ \hline
 0.2 & $0.311 \pm 0.001$ & $0.685 \pm 0.006$  & $\approx 10.35$  \\ \hline
 0.3 & $0.315 \pm 0.002$ & $0.744 \pm 0.006$  & $\approx 9.34$ \\ \hline
 0.4 & $0.333 \pm 0.004$ & $0.93 \pm 0.02$  & $\approx 8.95$ \\ \hline
 0.5 & $0.321 \pm 0.005$ & $1.09 \pm 0.01$ & $\approx 8.51$  \\ \hline
 0.6 & $0.317 \pm 0.004$ & $1.15 \pm 0.01$  & $\approx 8.44$ \\ \hline
 0.7 & $0.319 \pm 0.002$ & $1.24 \pm 0.02$  & $\approx 8.12$ \\ \hline
 0.9 & $0.311 \pm 0.005$ & $1.3 \pm 0.1$  & $\approx 7.78$ \\ \hline
 \end{tabular}

 \caption{Results of the extraction of parameters from the $\rho(\gamma)$ distribution shown in Fig. \ref{Fig3} (b) for $0.1 \leqslant H \leqslant 0.9 $. For $0.5\leqslant H\leqslant0.9$, a dominant exponential decay can be observed in the $\rho(\gamma)$ distribution, characterized by $\delta_{H}^{I}$ (see the text for more details).}
 \label{tab}
 \end{table}
%

\begin{figure}
\includegraphics [width=8.0cm] {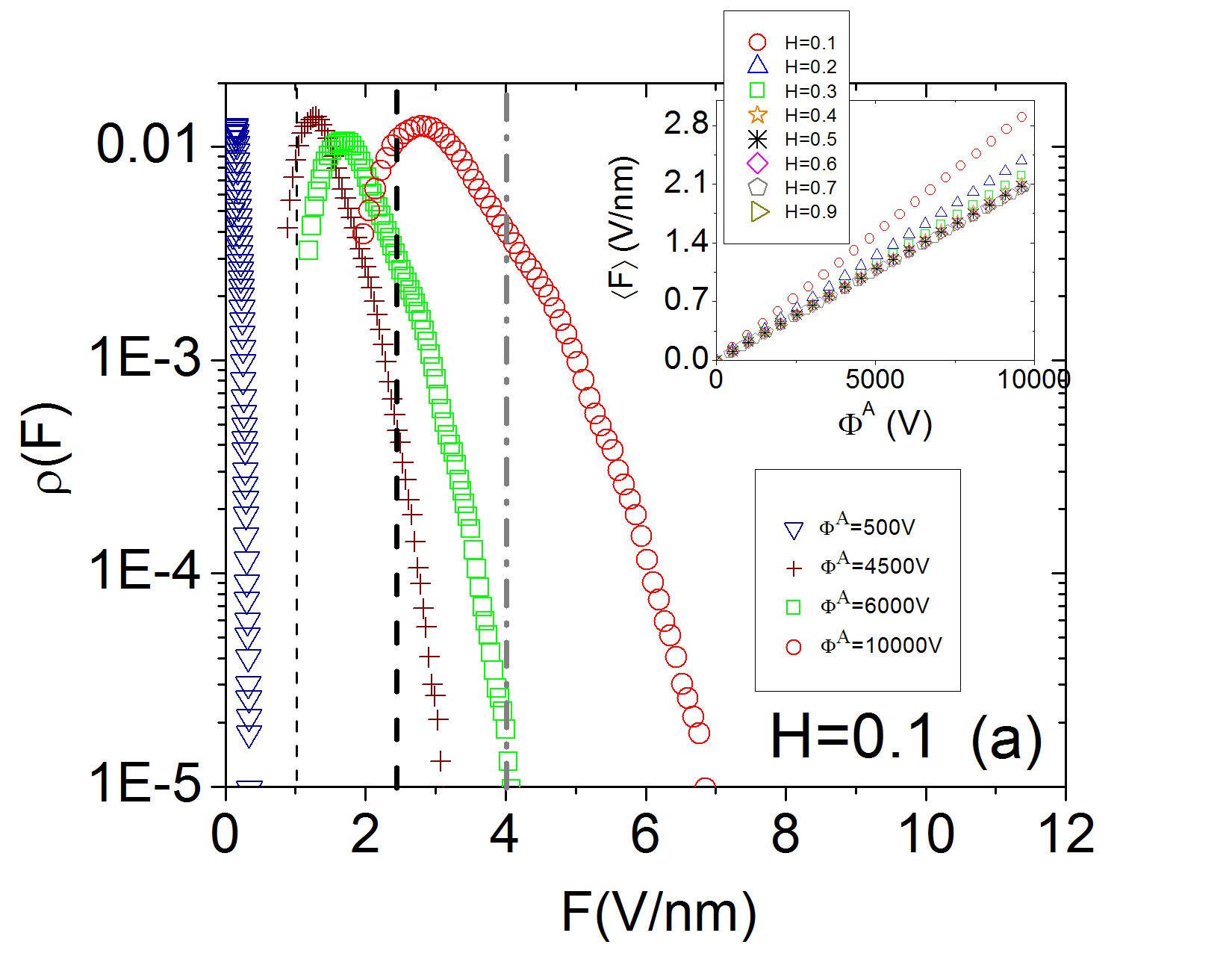}
\includegraphics [width=8.5cm] {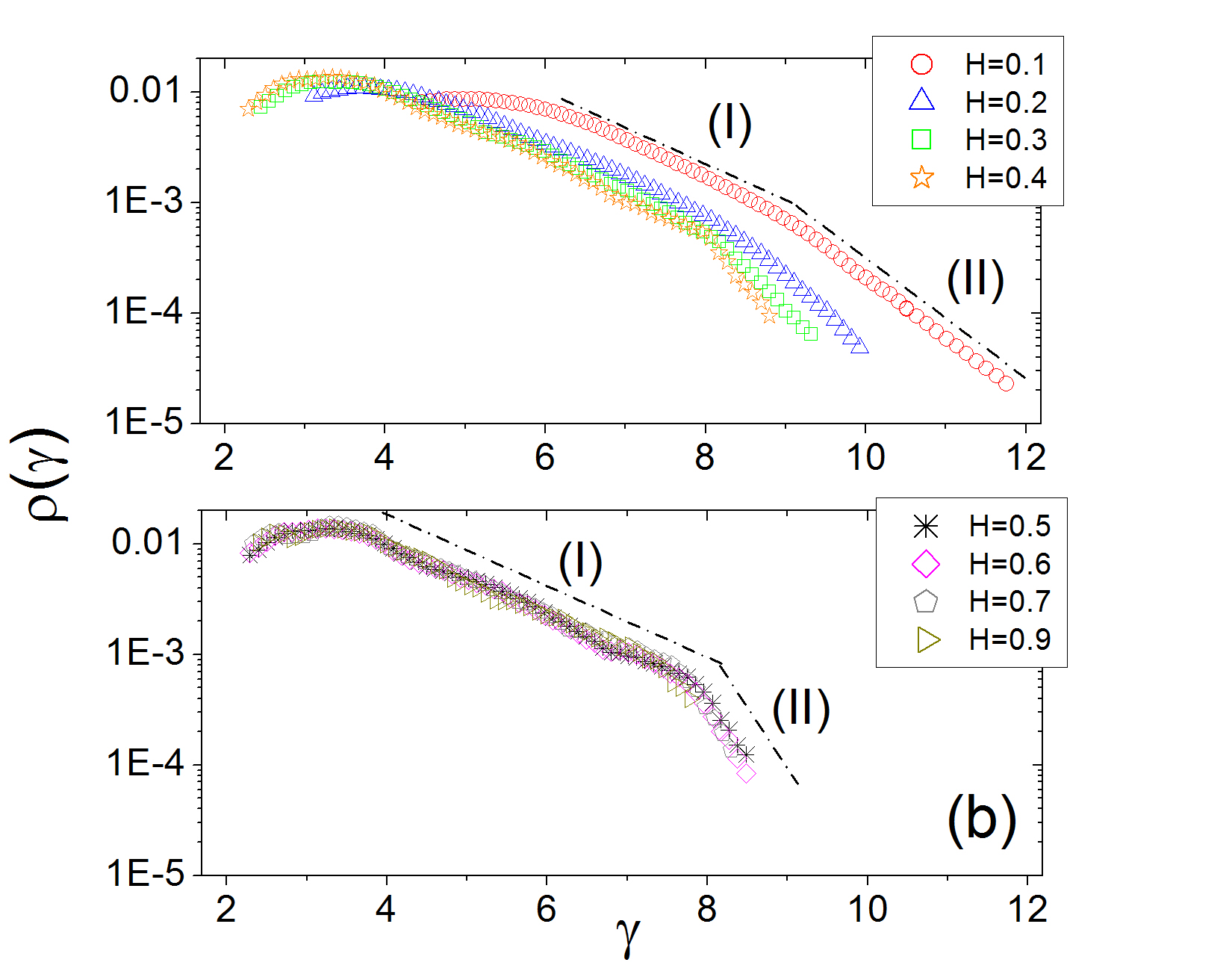}
\includegraphics [width=8.0cm] {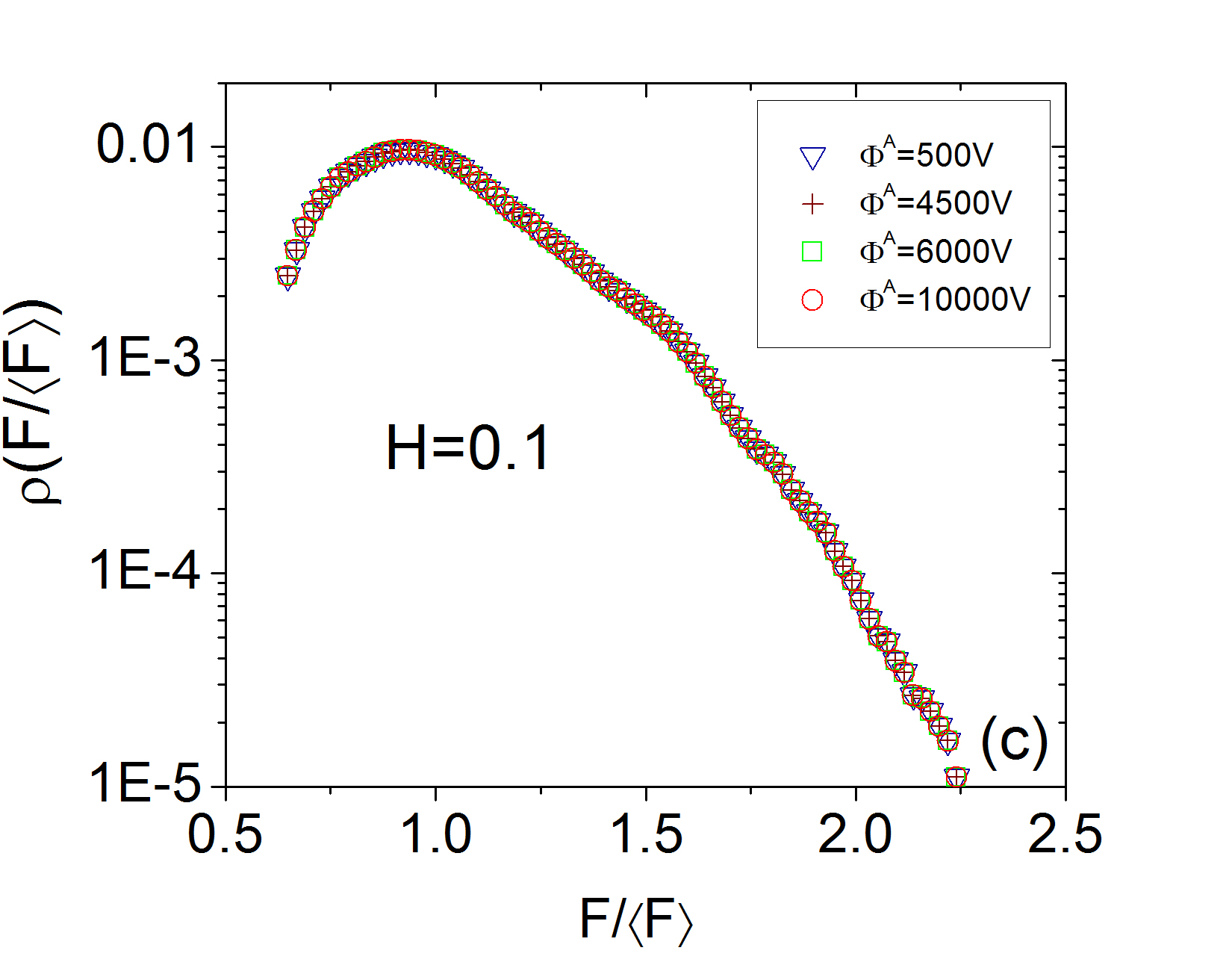}
\caption{(a) Local electric field intensity distribution, $\rho(F)$, for a LACFES with $H=0.1$. The vertical (black) dashed and (gray) dashed-dotted lines represent electric field intensities of $2.5$ V/nm and $4$ V/nm (see Ref. \cite{ForbesResub}), which correspond to typical values for field emission in pure metals with local work functions of approximately $3.5$ eV and $4.5$ eV, respectively. The inset shows the behavior of the mean electric field $\langle F \rangle$ (see Eq. (\ref{Eq2})) as a function of the anode electric potential $\Phi^{A}$ for $H=0.1$ (red circles), $H=0.2$ (blue triangles), $H=0.3$ (green squares), $H=0.5$ (black asterisks), $H=0.6$ (magenta diamonds), $H=0.7$ (gray pentagons) and $H=0.9$ (dark yellow right triangles). (b) Local FEF distribution, $\rho(\gamma)$, for $H \in [0.1,0.9]$. For a LACFES with $0.1 \leqslant H < 0.5$ (top panel), two pronounced exponentially decaying regimes, (I) and (II), can be clearly observed, and the dashed-dotted lines illustrate this behavior for $H=0.1$. For a LACFES with $0.5 \leqslant H \leqslant 0.9$ (bottom panel), it is evident that (I) and (II) correspond to slow and fast exponential decays, respectively, with the latter reflecting the low probability of finding high peaks on the corresponding LACFES. (c) Distributions of $\chi \equiv F/\langle F \rangle$ for $H=0.1$ for all $\Phi^{A}$ values considered in (a). The collapse of the curves is evident, indicating that the results for the $\rho(\gamma)$ distributions (for all $H$) do not depend on the anode electric potential $\Phi^{A}$.} \label{Fig3}
\end{figure}
Interestingly, for all $H$, two exponential decaying regimes are apparent, labeled as (I) and (II), which correspond to $\rho^{I}(\gamma) \sim \exp(-\delta^I_{H} \gamma)$ and $\rho^{II}(\gamma) \sim \exp(-\delta^{II}_{H} \gamma)$, respectively, where $\delta^{II}_{H} > \delta^I_{H}$, in general. In particular, for $H=0.1$, $\delta^{I}_{0.1} \times \log_{10} (e) = (0.343 \pm 0.001)$ is identified for interval (I) and $\delta^{II}_{0.1} \times \log_{10} (e) = (0.54 \pm 0.01)$ is found for interval (II) (the slopes of the dashed-dotted lines in Fig. \ref{Fig3}(b), where ``e" is Neper's number). In fact, $\Delta\delta \equiv \delta^{II}_{H} - \delta^{I}_{H}$ assumes the lowest value for $H=0.1$ (see Table I), suggesting that the area in which electron emission occurs is larger in the limit of large values of $\Phi^{A}$. This result, which is a consequence of the first (and second) slower decay of $\rho(\gamma)$, suggests a greater probability of finding emitting sites on LACFESs (primarily for $H=0.1$) with larger FEFs. We also compute the characteristic FEF, $\gamma_{C} \equiv F_{C}/F_{M}$, where $F_{C} \equiv max \{F_{i,j}\}$. For $H=0.1$, it is found that $\gamma_{C} \approx 11.86$. The same behavior of the two exponential regimes is observed for all $H$.

Table I summarizes the values of the parameters extracted from the $\rho(\gamma)$ distributions presented in Fig. \ref{Fig3} (b) and the $\gamma_C$ values for all values of $H$ explored in this work. Interestingly, these features capture important experimental results for thin-film emitters with irregular surfaces, in which $\rho(\gamma)$ has been shown to exhibit an exponential dependence on the FEF \cite{JAP}. However, we note that the FEF values reported in that work were typically between $100$ and $300$ (i.e., approximately one order of magnitude larger than those found in this work). This effect can be understood based on the following considerations: (i) the differences between the global roughness of the experimental surfaces used in the previous experiment and ours and (ii) the one-dimensional field variation across the emitter surface in our approach (see Supplementary Information II). In the cited work \cite{JAP}, the global roughness on a length scale of 20 $\mu$m was found to be approximately $W \approx 4 \mu$m, whereas in our work, the considered LACFESs have a roughness of $W \approx 75$ nm (i.e., $\sim 50$ times smaller) on the same length scale for all $H$ (see Fig. \ref{Fig1}).

In Supplementary Information II, we present strong evidence to corroborate speculation (ii) by simulating a genuine three-dimensional surface using a possible experimental setup modeled as an irregular emitting surface iteratively generated by a fractional Brownian motion ($FBM$) algorithm with $H=0.1$. In this case, again, an exponentially decaying regime in $\rho(\gamma)$ is observed. This finding indicates that our LACFES model captures the main features of the experimental results, namely, the local electric field distribution and the field emission properties (in the orthodox theory) of irregular conducting surfaces. This discussion suggests the importance of correctly measuring the $H$ exponents and the global roughnesses of the rough surfaces used to experimentally represent LACFESs (such as randomly oriented carbon thin-film emitters) in terms of, for instance, the local roughness scale (see Eq. (\ref{Eq1b})) extracted using a probe microscopy technique such as electrostatic force microscopy (EFM). In such a case, a careful analysis must be performed to account for the distortions in the image that may occur due to the finite size of the EFM tip \cite{EFM}, which may result in overestimation of the Hurst exponent (or the local roughness).
\\
\\
{\large \bf Area efficiency of emission and Fowler-Nordheim plots}
\\
\\
Previously, the occurrence of a relatively slow decay (interval (I)) was observed in the $\rho(\gamma)$ distribution for arbitrary $H$, as shown in Table (I). Remarkably, this tendency appears to be characterized by approximately the same parameter $\delta_{H}^{I}$ for any $H$ in this interval. This interesting result suggests that the area in which electron emission occurs in a LACFES may be higher compared with the ideal metal surface morphologies that are often modeled to explain experimental results. If a metal has no appreciable irregularities on the nanometer scale, this quantity corresponds to the Area Efficiency of Emission (AEE), $\alpha_{M}$, which is typically less than $10^{-5}$ (see Ref. \cite{Forbes3}).
This indicates that for a LACFES with a regular array of single emitters, it is expected that the effective emission area (the emitting area at the tips) will be considerably less than the apparent ``macroscopic" geometrical area (or ``substrate footprint" area $A_{M}$) of the physical emitter that is observed visually. Based on these arguments, if the emission area is, in fact, constant, it is convenient to define $\alpha_M$ as follows (here, $\alpha_M = \lambda_{M}$, using Forbes' notation (see Ref. \cite{Forbes3}), because we set $\lambda_{CFE}=1$; $\lambda_{M}$ is the macroscopic ``pre-exponential correction factor" or, alternatively, ``formal emission efficiency"):

\begin{equation}
\alpha_M \equiv \frac{J_{M}}{J_{kC}},
\label{Eq2a}
\end{equation}
where the \textit{characteristic kernel current density}, $J_{kC}$ (which is equal to the characteristic local current density, $J_{C}$, using Forbes' notation, by the same arguments presented above), and the \textit{macroscopic current density}, $J_{M}$, for a LACFES are given by

\begin{equation}
J_{kC} \equiv max\{J_{k}(\phi ,F)\},
\label{Eq3}
\end{equation}
where $J_{k}(\phi,F)$ is given by Eq. (\ref{Eq1f}) and, because $\lambda_{CFE}$ is approximated as equal to $1$,

\begin{equation}
J_{M} = i/A_{M} = \frac{\sum\limits_{p} J_{k}(z_{p}) \Delta A_{p}}{A_{M}}.
\label{Eq4}
\end{equation}
Here, ``$i$" is the total emission current, and the sum is taken over the ``substrate footprint" area of the emitter, $A_M$. Furthermore, the values of $J_{k}(z_{p})$ are given by Eq. (\ref{Eq1f}). These assumptions are equivalent to assuming that the empirical CFE (i,$\Phi^{A}$) characteristics should obey \cite{ForbesAPL}

\begin{equation}
i = C (\Phi^{A})^{\kappa} \exp{(-N/\Phi^{A})},
\label{Eq4a}
\end{equation}
where $C$, $N$ and $\kappa$ are constants. If the electron field emission is orthodox, considering the SN barrier, and the emission area is constant, then $\kappa$ is expected to be $\kappa = 2 - \eta(\phi)/6$, where $\eta(\phi)$ is given by Eq. (\ref{Eq1d3}) (see Ref. [32]). In Eq. (\ref{Eq4a}), we assume that the emission quantities and the measured quantities $(i,\Phi^{A})$ are identical.

Note that the linear dependence between $J_{M}$ and $J_{kC}$ for a rough LACFES is not clear in our case because the LACFES exhibits irregularities on small scales (with the apex radii of the emitters greater than dozens of nanometers). This question, for the geometries used in this work, has not been previously addressed; therefore, we must address it. Studies that consider smooth surfaces generally follow the implicit assumption that $\alpha_M$ is only weakly field dependent, meaning that for practical purposes, one can take it to be nearly constant. However, we show that this assumption does not hold, particularly for LACFESs with high local roughness ($ 0.1 \leqslant H < 0.5$). As we will discuss, the consequences of the scale relation between $J_M$ and $J_{kC}$ can provide relevant information concerning the effects of the morphologies of rough experimental LACFESs on the FN plot slopes if the electron field emission is orthodox.

\begin{figure}
\includegraphics [width=9.5cm] {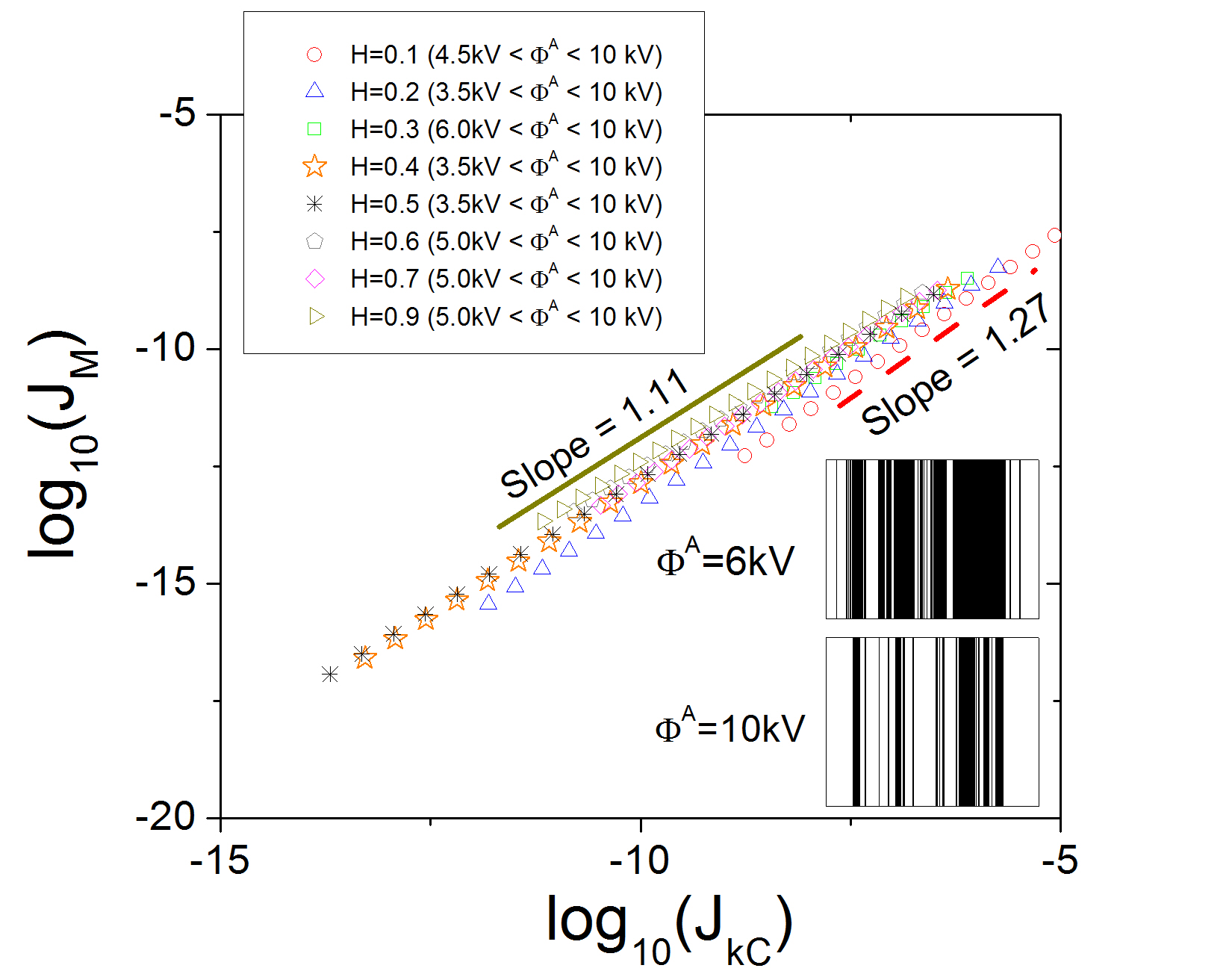}
\caption{Macroscopic current density, $J_{M}$ (see Eq. (\ref{Eq4})), as a function of the characteristic kernel current density, $J_{kC}$ (see Eq.(\ref{Eq3})), for $0.1 \leqslant H \leqslant 0.9$ and for an anode potential in the range $3.5 kV < \Phi^{A} < 10 kV$. The slopes (values of $\beta_{H}$ - see Eq.(\ref{Eq6a})) for $H=0.1$ and $H=0.9$ were found to be $1.27$ (dashed line) and $1.11$ (solid line), respectively, with an error of $10^{-3}$ in the linear regression fit (see Table II for other values of $\beta_{H}$). The inset shows a snapshot of a portion of a LACFES with $H=0.1$, in which the non-emitting locations (black) are distinguished from the emitting locations (white). The substantial increase in the effective area (see the text for more details) as the anode electric potential changes from $\Phi^{A}=6 kV$ to $\Phi^{A}=10 kV$ can be clearly observed.} \label{Fig3f}
\end{figure}

We emphasize that we use an approximation in which each point on the cathode surface represents a particular lateral location on the emitter surface, as previously discussed. Moreover, in our calculations, space is filled with a discretized cubic lattice with a unit volume of $\Delta x \times \Delta y \times \Delta z$. Thus, Eq. (\ref{Eq4}) can be written as

\begin{equation}
J_{M} \approx \frac{\sum\limits_{p} J_{L}(z_{p}) \Delta x_{p} \Delta y_{p} }{(N_{x}-1) \Delta x \times (N_{y}-1)\Delta y},
\label{Eq5}
\end{equation}
where $N_{x}$ and $N_{y}$ represent the numbers of points defining the LACFESs in the $x$ and $y$ directions, respectively.
In Eq.(\ref{Eq5}), $\Delta x_{1}
\Delta y_{1} =...=\Delta x_{N_{x}}\Delta y_{N_{y}} = \Delta x\Delta y$.

For LACFESs (for all $H$), we follow the definition of Eq. (\ref{Eq3}) by identifying ``C" with the apexes of the more prominent emission sites. In Eq. (\ref{Eq4}), if we use the SN barrier of unreduced height $\phi$ and the Forbes approximation \cite{Forbes}, then $\nu_{F} = \nu(\textit{f}) \approx 1-\textit{f}+(1/6)\textit{f}\ln(\textit{f})$, with $\textit{f}$ given by Eq. (\ref{Eq1d2}). We assume that for the LACFESs considered in this work, the local work function is approximately constant over the surface, and we adopt $\phi = 3.5$ eV. Therefore, the parameter given by Eq. (\ref{Eq1d3}) is considered to be $\eta \approx 5.26$.

Thus, by introducing the Forbes approximation for the SN barrier correction function $\nu(\textit{f})$, the exponential factor in Eq. (\ref{Eq1f}) can be expanded as follows:

\begin{equation}
\exp\left[-\nu(\textit{f}) b \phi^{3/2}/F\right] \approx
  e^\eta \; \left[\frac{F}{F_{R}}\right]^{-\eta/6} \;
  \exp\left[-b \phi^{3/2}/F\right].
\label{Eq6}
\end{equation}
Fig. \ref{Fig3f} shows the behavior of $\log_{10}(J_{M})$ as a function of $\log_{10}(J_{kC})$ for $0.1 \leq H \leq 0.9$ and an anode potential in the range $3.5 kV < \Phi^{A} < 10 kV$. These results suggest a scaling relation between $J_{M}$ and $J_{kC}$ as follows:

\begin{equation}
J_{M} \sim \left[J_{kC}\right]^{\beta_H} \hspace{1.0cm} (\beta_{H}>1),
\label{Eq6a}
\end{equation}
where the exponent $\beta_{H}$ depends on the $H$ of the LACFES. According to Eq. $\ref{Eq2a}$, this result can be understood as the result of a power-law dependence between $\alpha_M$ (equal to $\lambda_M$ in our case) and $J_{kC}$, i.e., $\alpha_M = c_1 (J_{kC})^{c_2}$, where, $c_1 > 0$ and $c_2 > 0$ ($c_1 = \alpha_M$ and $c_2$=0 if the emission area is constant) are constants in an appropriate range of the macroscopic electric fields. Indeed, if the surface of a LACFES exhibits fluctuations on small scales, then for a low electric potential of the anode (or, alternatively, a macroscopic electric field, $F_M$), only some points on the surface (those with larger FEF values) are emitters. If $F_M$ increases (decreases) by a factor $\Delta > 1$ ($0 < \Delta < 1$), $J_{kC}$ also increases (decreases) by a factor $\Delta^{2-\eta/6} \left[ \exp\left(-b \phi^{3/2}/(\gamma_{C}F_{M})\right) \right]^{{(1/\Delta} - 1)}$. Moreover, some new locations on the surface will become emitters in addition to those that were previously emitters. Thus, $\alpha_{M}$ increases (decreases) by a factor $\Delta^{(2-\eta/6)c_{2}} \left[ \exp\left(- c_{2} b \phi^{3/2}/(\gamma_{C}F_{M})\right) \right]^{{(1/\Delta} - 1)}$. This is a very reasonable physical explanation because of the relatively slow decay which may be observed in the first and second regions of the local electric field distributions (for instance, corresponding to $H=0.1$), which indicates that there are large regions on a LACFES where the differences in FEF values are relatively small. Thus, a small variation in the macroscopic electric field can result in the appearance of a non-linear relation between $\alpha_M$ and $J_{kC}$. If the fluctuations of the FEFs on the LACFES are sufficiently large, then the emission locations with smaller FEF values may not be able to become field emitter, even at high experimental electric fields. The role of the exponent $\beta_{H}$ ($\equiv 1 + c_2$) is reflected in this behavior.

It is evident that the linear dependence between $J_M$ and $J_{kC}$ (as a consequence of the lack of dependence between $\alpha_{M}$ and $J_{kC}$, as shown in Eq. (\ref{Eq2a})) is more approximate for $0.5\leqslant H \leqslant0.9$, where the height fluctuations of the emitting surface are correlated (whereas they are random for $H=0.5$). Outside this range, the presence of anticorrelations and, consequently, more small-scale irregularities of the LACFES contribute more strongly to the increase in the CFE area. The numerical calculations corroborate this assumption, and the corresponding values of $\beta_H$ can be determined from the slopes of the curves shown in Fig. \ref{Fig3f}. The results indicate that for $0.1 \leqslant H< 0.5$ in particular, $\beta_{H}$ is significantly different from unity. The slopes for $0.1 \leqslant H \leqslant 0.9$ are presented in Table II. This observation indicates that for rough LACFESs, small Hurst exponents play an important role in causing the rapid increase (relative to that observed in the case of smoother LACFESs) in the ``effective area" across the surface of the field emitter as the macroscopic electric field $F_M$ (or, alternatively, $\Phi^{A}$) increases. This may be a signature of an interesting scale-invariant relation between $J_{M}$ and $J_{kC}$ (as the result of a power-law dependence between $\alpha_M$ and $J_{kC}$), which can motivate experimental tests with real morphologies that follow orthodox field emission.

The inset of Fig. \ref{Fig3f} shows a representation of a portion of a LACFES with $H=0.1$, which illustrates the contrast between the non-emitting and emitting locations. To define these features, we apply a criterion such that a given location on the LACFES is an emitter if $J_{k}(\phi,F)/J_M > 10^{-3}$. The substantial increase in the effective area that occurs as the electric potential of the anode changes from $\Phi^{A}=6 kV$ to $\Phi^{A}=10 kV$ is evident. This result is also a consequence of the two exponential regimes with small values of $\Delta\delta$ (see Table I) in the distribution $\rho(\gamma)$, which is more pronounced for $0.1\leqslant H<0.5$, thereby increasing the probability of finding emitter locations with high FEF values.

Now we address our primary purpose, namely, elucidating the effect of the Hurst exponent of an LACFES on the slope of the FN plot, which is typically used to extract relevant features of field emitter surfaces. Experimentally, the data obtained from field emission measurements can be described as orthodox when the characteristic field enhancement factor, $\gamma_{C}$, is independent of voltage. Indeed, in our model, this requirement is satisfied. Moreover, we assume that the emission is controlled solely by the tunneling barrier at the emitter/vacuum interface. Thus, our results can be tested experimentally under these conditions.
Fig. \ref{Fig4} shows the behavior of $\ln\{(J_{M}/(F_M)^{2})\}$ vs. $1/F_{M}$ (or a $J_{M}$-$F_{M}$-type FN plot) for a LACFES with $0.1\leqslant H \leqslant0.9$. The FN plot appears to exhibit approximately linear behavior in the considered range of $\Phi^{A}$, although in reality, the derivative is not constant, as observed in the inset of Fig. \ref{Fig4} for $H=0.1$ and $H=0.9$. An approximately constant value of the derivative is observed in the limit of low macroscopic electric fields, in which only the main peaks are emitters, followed by an increase in the corresponding absolute value in the limit of high values of $F_{M}$.

The slope of the $J_{M}$-$F_{M}$-type FN plot (extracted from a linear least-squares fit to the numerical (or experimental) data), which depends on $H$, is denoted by $S_{F_M}(H)$. In fact, the higher is the characteristic FEF value, the lower in magnitude the slope of the FN plot will be. This trend is evident in Fig. \ref{Fig4} and reflects the effect of the different Hurst exponents of the LACFESs (for the same global roughness ($W \thickapprox 75$ nm) - $S_{F_M}(0.5) > S_{F_M}(0.3)> S_{F_M}(0.2) > S_{F_M}(0.1)$), thereby corroborating the inequality in the corresponding $\gamma_{C}$ values shown in Table I. The values of $S_{F_M}(H)$ are presented in Table II.

\begin{figure}
\includegraphics [width=9.5cm] {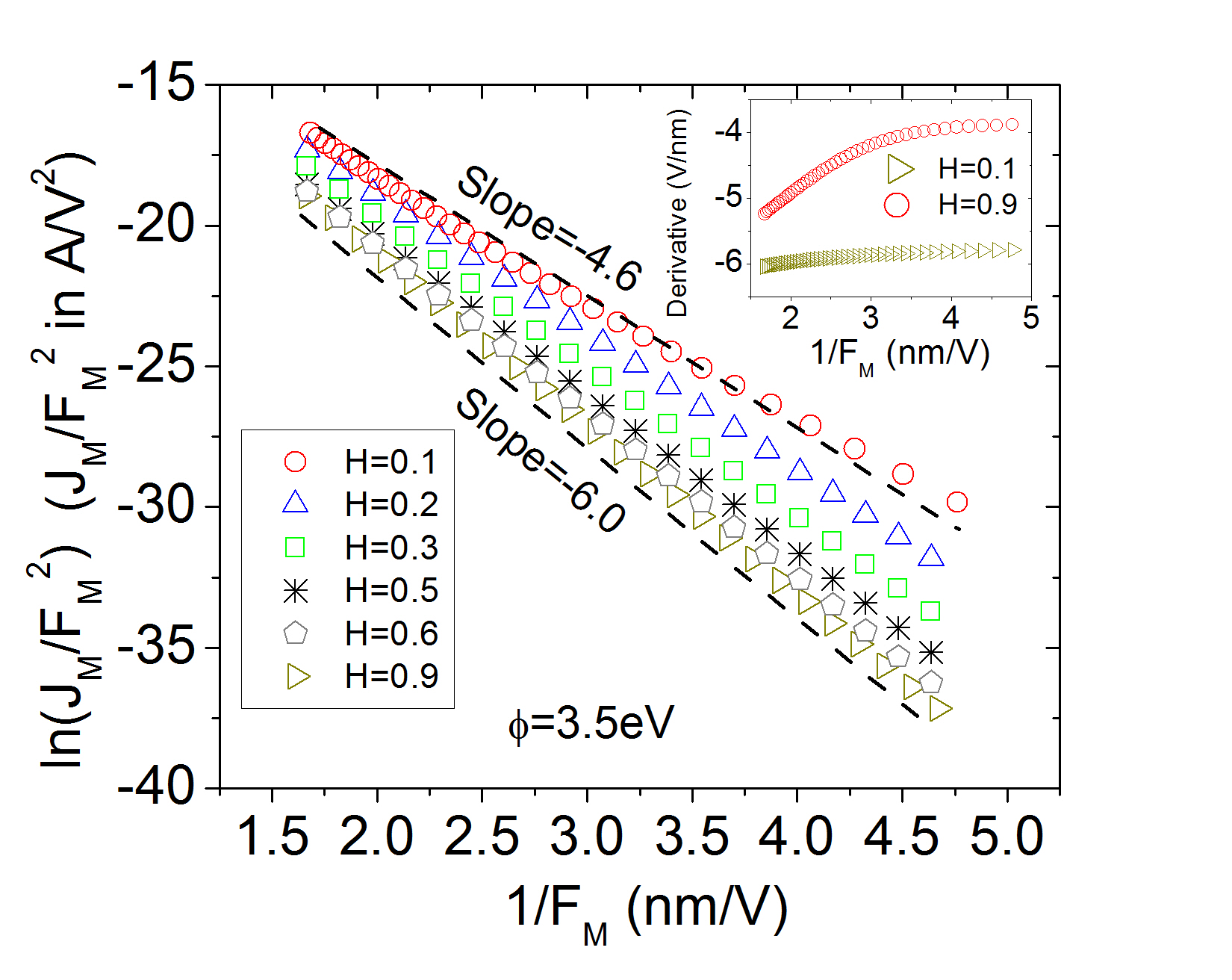}
\caption{$J_{M}$-$F_{M}$-type FN plot for LACFESs with $0.1\leq H\leq0.9$. The corresponding slopes for $H=0.1$ and $H=0.9$ (dashed lines) are shown (see Table II for other values of $S_{F_M}(H)$ corresponding to $0.1 \leqslant H \leqslant 0.9$). The work function of the LACFES is considered to be approximately constant and equal to $\phi = 3.5$ eV. The inset of this figure shows the derivatives of the $J_{M}$-$F_{M}$-type FN plots for LACFESs with Hurst exponents of $H=0.1$ and $H=0.9$.} \label{Fig4}
\end{figure}
First, we use the elementary slope characterization parameter (SCP), which is denoted by $\gamma_{C}^{T}$ (see Ref.[4]), to simulate the common practice of experimentalists in using the triangular-barrier (or elementary) FN equation (Eqs. (\ref{Eq1e}) and (\ref{Eq1f}) for $\lambda_{CFE} = \nu_F = 1$). In this case, the SCP is given by

\begin{equation}
\gamma_{C}^{T} \equiv -b \phi^{3/2} / S_{F_M}(H).
\label{Eq8}
\end{equation}
From the slopes presented in Fig. \ref{Fig4} (see Table II) and using Eq. (\ref{Eq8}), we calculate $\gamma_{C}^{T}= 7.37$ (error of 5\% compared with the $\gamma_C$ value reported in Table I) for $H=0.9$ and $\gamma_{C}^{T} = 9.72$ (error of $\approx 20$\% compared with the $\gamma_C$ value reported in Table I) for $H=0.1$. This latter error, which is related to the fact that there is a greater probability of finding field emitter locations on the LACFES with larger FEF values than is naively expected (as is also the case for $H$ values in the interval $0.1<H<0.4$), represents an underestimation of the real characteristic FEF value that is of practical significance. Table II summarizes the values of $\gamma_{C}^{T}$ calculated for $0.1 \leqslant H \leqslant0.9$. To further explain this result, let us return to the form of the scaling between $J_M$ and $J_{kC}$. From Eqs. (\ref{Eq1f}) and (\ref{Eq6a}), it follows that

\begin{equation}
\ln\{(J_{M}/(F_M)^{2})\} = \Gamma\left(\frac{\partial \left(\ln[J_{M}(H)/F_{M}^{2}]\right)}{\partial \left(1/F_M\right)}, F_{M}^{2(\beta_{H} - 1)}\right)
- \left(\frac{\beta_H \nu_{F_{C}} b \phi^{3/2}}{\gamma^{\beta_H,\sigma}_{C}}\right) \times \frac{1}{F_{M}},
\label{Eq9}
\end{equation}
from which it is possible to define the corrected FEF:

\begin{equation}
\gamma^{\beta_H,\sigma}_{C} \equiv -\left(\frac{\beta_H \sigma b \phi^{3/2}}{S_{F_M}(H)}\right).
\label{Eq10}
\end{equation}
In Eq. (\ref{Eq10}), $\sigma$ is a generalized slope correction factor related to the SN barrier and $\beta_H$ includes the effect of the geometry of the LACFES on the estimation of the characteristic FEF. Formally, the first term in Eq. (\ref{Eq9}) does not define the intercept in the FN plot. This occurs only if a triangular barrier, which corresponds to an unrealistic physical situation, is considered and if the area of emission of the rough LACFES is independent of voltage. This is not the case in our system, where the effects that can produce non-linear features in the $J_{M}$-$F_{M}$-type FN plot are the dependence between the area of emission of the rough LACFES and the anode voltage as well as the effect of the SN barrier. According to this argument, the macroscopic current density is a function of $H$ such that $J_{M}=J_{M}(H)$. The process of deriving emission-area estimates from FN plots has previously been investigated by Forbes \textit{et al.} \cite{ForbesArea} using free-electron theory and considering three different tunneling-barrier models. However, the current work can motivate the investigation of an additional effect, namely, that of the irregular morphology of the LACFES. In this way, it is possible to define, based on the function $\Gamma$ in Eq. (\ref{Eq9}), an effective Area Efficiency of Emission, $\alpha_{eff}$, that clearly depends on both the macroscopic electric field, $F_{M}$, and the derivative $\frac{\partial \left(\ln[J_{M}(H)/F_{M}^{2}]\right)}{\partial \left(1/F_M\right)}$. This quantity, $\alpha_{eff}$, can be extracted from the term $\Gamma\left(\frac{\partial \left(\ln[J_{M}(H)/F_{M}^{2}]\right)}{\partial \left(1/F_M\right)}, F_{M}^{2(\beta_{H} - 1)}\right)$ in a given range of values of the variable $1/F_{M}$, again because most emitters melt for values of $1/F_{M}$ less than some reference value \cite{ForbesSlope} (and this effect may be more pronounced in the case of a rough LACFES).

\begin{table}
   \centering
 \renewcommand{\arraystretch}{1.5}


 \begin{tabular}{|c|c|c|c|c|}

  \hline
 $H$ & $S_{F_M}(H) (V/nm)$ & $\gamma_{C}^{T}$ & $\beta_H$ & $\gamma^{\beta_H,\sigma}_{C}$  \\

 \hline

 0.1 & $-4.60 \pm 0.05$ & $9.72 \pm 0.05$ & $1.271 \pm 0.001$ & $11.79 \pm 0.05$    \\ \hline
 0.2 & $-4.88 \pm 0.03$ & $9.16 \pm 0.03$ & $1.185 \pm 0.001$ & $10.32 \pm 0.03$   \\ \hline
 0.3 & $-5.32 \pm 0.02$ & $8.40 \pm 0.01$ & $1.174 \pm 0.001$ & $9.39 \pm 0.01$ \\ \hline
 0.4 & $-5.45 \pm 0.02$ & $8.20 \pm 0.01$ & $1.132 \pm 0.001$ & $8.87 \pm 0.01$ \\ \hline
 0.5 & $-5.60 \pm 0.01$ & $7.98 \pm 0.01$ & $1.127 \pm 0.001$ & $8.59 \pm 0.01$ \\ \hline
 0.6 & $-5.87 \pm 0.02$ & $7.61 \pm 0.02$ & $1.145 \pm 0.001$ & $8.33 \pm 0.02$ \\ \hline
 0.7 & $-5.92 \pm 0.02$ & $7.55 \pm 0.02$ & $1.124 \pm 0.001$ & $8.10 \pm 0.02$ \\ \hline
 0.9 & $-6.06 \pm 0.01$ & $7.37 \pm 0.01$ & $1.113 \pm 0.001$ & $7.82 \pm 0.01$ \\ \hline
 \end{tabular}

\caption{Results for $0.1\leqslant H \leqslant 0.9$. The slopes of the $J_{M}$-$F_{M}$-type FN plots shown in Fig. \ref{Fig4}, the characteristic FEF values $\gamma_{C}^{T}$, the $\beta_{H}$ values extracted from the results presented in Fig.\ref{Fig3f} and $\gamma^{\beta_H,\sigma}_{C}$. $\gamma_{C}^{T}$ and $\gamma^{\beta_H,\sigma}_{C}$ were calculated using Eqs. (\ref{Eq8}) and (\ref{Eq10}) while considering the elementary FN equation and relevant corrections, including the SN barrier and the morphology of the LACFES, respectively.}
\label{tab}
\end{table}

Obviously, because $\beta_{H}$ is close to unity ($F_{M}^{2(\beta_{H} - 1)} \rightarrow 1 $), $\alpha_{eff} \approx \alpha_{M}$. Interestingly, this feature also appears, more approximately, in the case of $0.5 \leqslant H \leqslant 0.9$, in which the height fluctuations on the LACFES are statistically correlated (or random). $\sigma$ typically takes a mid-operating-range value of approximately $0.955$ \cite{Forbes3,ForbesSlope}. If we apply this assumption (which is certainly a good approximation for our system), consider the values of the FN plot slopes obtained from Fig. \ref{Fig4} (see Table II) and use Eq. (\ref{Eq10}) to estimate the corrected characteristic FEF value, $\gamma^{\beta_H,\sigma}_{C}$, then we obtain the values presented in Table II. Impressive agreement is clearly evident between $\gamma^{\beta_H,\sigma}_{C}$ and the values presented in Table I that were extracted from the distribution $\rho(\gamma)$, which is not accessible to field emission experimentalists. Thus, the correction $\beta_H$ must be applied (primarily for LACFESs with large amounts of individual field emitter tips) to more precisely extract the characteristic FEF of LACFESs from FN plots. The effect discussed above can, of course, alternatively be described as being caused by the variation in the effective emission area with voltage (or, equivalently, with the macroscopic field), and our conclusion is that this variation must be taken into account.

Alternatively, when empirical CFE (i-V) characteristics are used (where ``V" is the measured voltage) \cite{ForbesAPL}, our results suggest replacing $\kappa = 2 - \eta(\phi)/6$ (the result from Eq. (\ref{Eq4a}), considering the assumptions that lead to Eqs. (\ref{Eq2a}) and (\ref{Eq3}) as well as the expansion (\ref{Eq6})) with $\kappa(H) = (2 - \eta(\phi)/6)\beta_{H}$ for a rough LACFES (for constant $\beta_{H}$), if the range of the macroscopic electric field yields an ``adequately linear" FN plot. This suggests that an investigation of the real irregular morphology of the LACFES, through the precise measurement of the Hurst exponent (and the global roughness), is an important method of gaining a more complete understanding of the relevant measurements to be extracted from the FN plot to evaluate the electron emission capability of a material with an irregular morphology. Moreover, a systematic and careful investigation of the scaling between $J_M$ and $J_{kC}$, which are both experimentally accessible measurements, is of fundamental importance for evaluating the effects of the rough geometry of the LACFES on the corresponding CFE properties. Situations in which $\alpha_M$ scales as a more complicated function of $J_{kC}$ cannot be disregarded, and an investigation of these features in potential CFE materials is certainly desirable. Finally, our model can also be applied to investigate the inclusion of work function distributions over rough LACFESs, which has not been considered here. The investigation of this effect will be a subject of future work.

In summary, we investigated the role of the Hurst exponents of rough LACFESs on the corresponding electric field distributions and the field emission quantities of technological interest, particularly the characteristic FEF value, $\gamma_C$. Our results, which were obtained by considering the orthodox CFE and LACFESs with a global roughness on the order of a few tens of nanometers, demonstrated that correlated (or random) LACFES morphologies ($0.5 \leqslant H \leqslant 0.9$) exhibit a more weak dependence between the area of emission and the macroscopic electric field, whereas for $0.1 \leqslant H < 0.5$, a stronger dependence is evident that contributes to the formation of non-linear features in the FN plot. This is a consequence of small-scale fluctuations in the morphology of the field emitters, even when all LACFESs have the same global roughness.
For any $H$, the local electric field distribution over the LACFES exhibits two exponential regimes and a power-law scaling between the macroscopic current density ($J_M$) and the characteristic kernel current density ($J_{kC}$) for the typical experimental range of macroscopic electric fields. Our results indicate a scaling of the form $J_{M} \sim \left[J_{kC}\right]^{\beta_{H}}$, where the exponent $\beta_{H} > 1$ depends on $H$. Moreover, for orthodox field emission from a LACFES, this scale must be considered when calculating the slope characterization parameter (in addition to the generalized slope correction from the SN barrier) that is used by experimentalists to extract the characteristic FEF. Failing to include this parameter in the SCP (using FN plot data) may result, particularly for LACFESs with $0.1\leqslant H<0.5$, in a significant underestimation of the characteristic FEF, which is a very useful measure of the emitter sharpness.
\\
\\
{\large \bf Acknowledgements}
\\
\\
The author thanks R. G. Forbes for fruitful discussions and valuable suggestions to use the orthodox CFE theory in rough LACFES.
This work has been supported by CNPq (Brazil) under grant 150874/2014-6.

*

\vspace{2.5cm}

\maketitle

{\large \bf Supplementary Information I}
\\
\\
{\large \bf Numerical method for electric field calculations}
\\
\\
Assuming that the region between the emitter surface and the anode is vacuum,
this implies the numerical solution of Laplace's equation,
\bigskip
%
\begin{equation}
 \nabla^2 \Phi = 0,
  \label{EqA1}
\end{equation}
\bigskip
in a discretized space, imposing suitable boundary conditions.
In our case, they correspond to Dirichlet conditions both at
the cathode ($\Phi=0$) and the anode ($\Phi=\Phi^{\rm A}$),
and periodic lateral conditions,
so that for a domain formed by $L_{x} \times L_{y} \times L_{z}$ points
(in the directions $x,y,z$, respectively), we have
\bigskip
%
\begin{equation}
 \Phi_{i,j,k}=\Phi_{i+L_{x},j,k}=\Phi_{i,j+L_{y},k}=\Phi_{i+L_{x},j+L_{y},k}.
   \label{EqA2}
\end{equation}
\bigskip

Equation (\ref{EqA1}) is then solved iteratively using a second order
finite difference scheme,
in which the electric potential is given by
\bigskip
\begin{equation}
 \Phi_{i,j,k}^{(m+1)} = \frac{1}{6} [ \Phi_{i-\omega,j,k}^{(m)}
     \Phi_{i+\omega,j,k}^{(m)} + \Phi_{i,j-\omega,k}^{(m)} + \Phi_{i,j+\omega,k}^{(m)} +
     + \Phi_{i,j,k-\omega}^{(m)}
     + \Phi_{i,j,k+\omega}^{(m)} ].
  \label{EqA3}
\end{equation}
That is, at each iteration the new potential $\Phi^{(m+1)}_{i,j,k}$
is just the average of its magnitude on the surrounding points at the previous iteration, $(m)$. These values are then refined iteratively starting from the first step,
$\Phi_{i,j,k}^{(0)}=0$, until a satisfactory convergence criterion
\bigskip
\begin{equation}
 \epsilon_{\rm max} = \max \{\epsilon_{i,j,k}\} < \xi ,
   \label{EqA4}
\end{equation}
\bigskip

where
\bigskip
%
\begin{equation}
\epsilon_{i,j,k} = \left| \frac{ \Phi^{(m+1)}_{i,j,k}
 - \Phi^{(m)}_{i,j,k}}{\Phi^{(m+1)}_{i,j,k}} \right|,
\label{EqA5}
\end{equation}
\bigskip
is met. In this work, the error is set to be $\xi=10^{-6}$. In Eq.(\ref{EqA3}), $\omega$ is the length of lattice parameter in the cubic lattice. In this work, we refine
the grid in such that $\omega = 1/50u$, where $u$ is the basic unit distance.
However, no significatively deviations has been detected using $\omega = u$. Once convergence has been achieved, the intensity of the electric field,
\bigskip
%
\begin{equation}
\mathbf{F}(\mathbf{r}) = - \mathbf{\nabla}\Phi(\mathbf{r}),
\label{EqA6}
\end{equation}
\bigskip
at a any point (not necessarily on the grid), $\mathbf{r}_P=(x_P,y_P,z_P)$,
can be evaluated by linear interpolation from the values
at the eight vertices of the circumscribing grid cube.
If such a cube ``starts'' at the grid location ($i,j,k$),
the corresponding cartesian components, $F^{x,y,z}(\mathbf{r})$,
are given by
\bigskip
%
\begin{equation}
    F^\mu (r) =  N \sum_{\alpha_i,\alpha_j,\alpha_k=0}^\omega
    \frac{E^\mu_{i+\alpha_i,j+\alpha_j,k+\alpha_k}}
    {r(x_{P,\alpha_i},y_{P,\alpha_j},
    z_{P,\alpha_k})},
     \label{EqA7}
\end{equation}
%

where:
\begin{equation}
  \frac{1}{N} = \sum_{\alpha_i,\alpha_j,\alpha_k=0}^\omega
    \frac{1}{r(x_{P,\alpha_i},y_{P,\alpha_j},
    z_{P,\alpha_k})},
  \label{EqA8}
\end{equation}
\bigskip
and

\begin{equation}
r(x_{P,\alpha_i},y_{P,\alpha_j},
    z_{P,\alpha_k}) \equiv \\
    \sqrt{(x_P-x_{i+\alpha_i})^2+(y_P-y_{j+\alpha_j})^2+
    (z_P-z_{k+\alpha_k})^2}.
\end{equation}

\begin{figure}
\includegraphics [width=8.5cm] {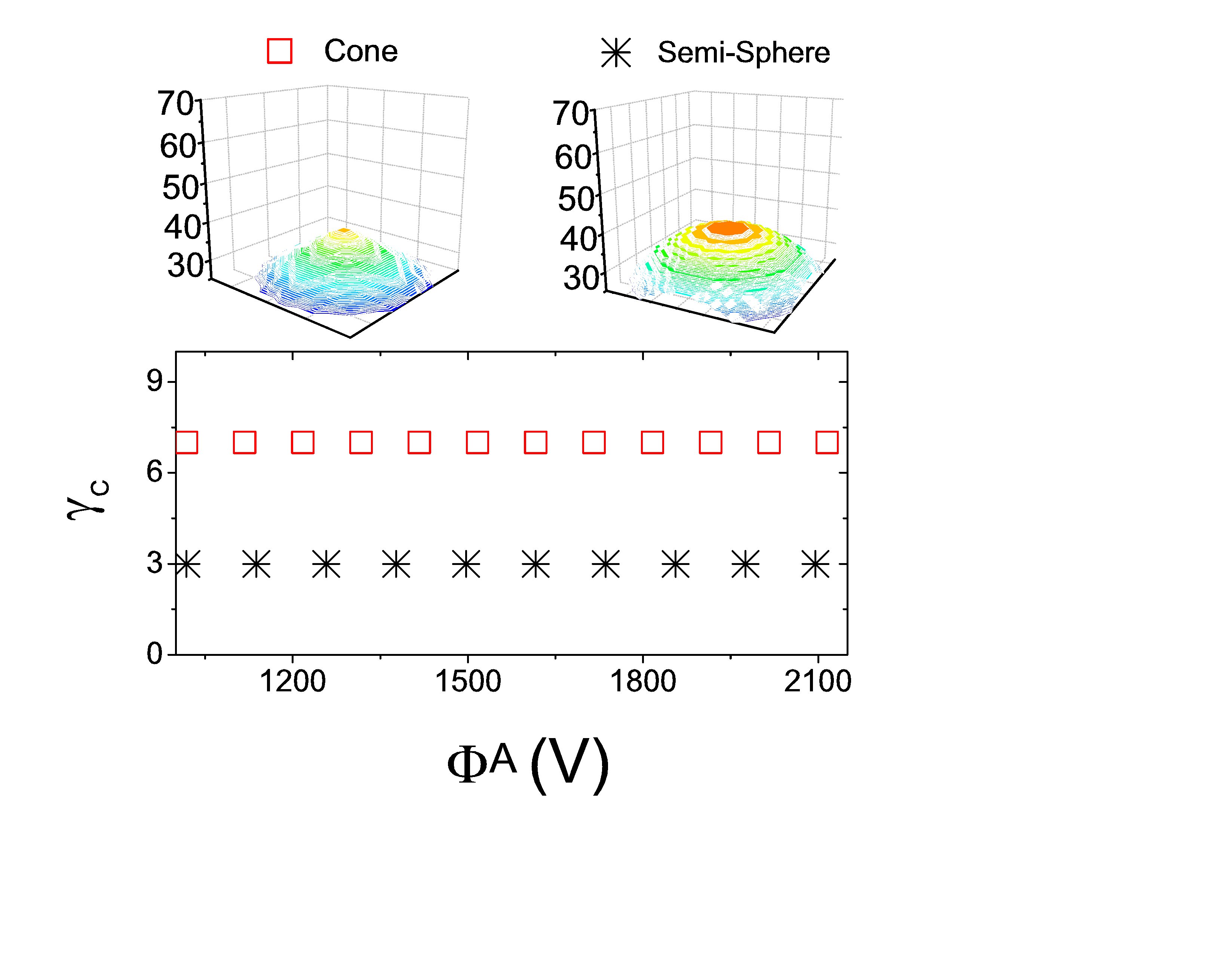}
\caption{Characteristic FEF, evaluated numerically by the solution of the Laplace equation, as a function of the anode electric potential, $\Phi^{A}$, considering a semi-sphere and a cone protuberances superimposed to a planar conducting surface.} \label{FEFSC}
\end{figure}

To test the accuracy of our method, we consider calculate the
characteristic FEF, $\gamma_C$, solving the electric field in three dimensions
around a conducting sphere (and conical) structure superimposed
in the center of a planar conducting substrate.
The electric potential at the surface of the conducting sphere was
fixed to $\Phi=0$, while at the anode a electric potential $\Phi^{A}$.
The results for $\gamma_C$ of the two structures are presented in Fig. \ref{FEFSC}, for $\Delta x_{i},\Delta x_{j} \ll L$,
as a function of the electric potential $\Phi^{A}$. It's clear the very good agreement between our numerical results and
the analytical solution which predicts $\gamma_C = 3$ for the semi-sphere on a plane \cite{ForbesFEF}. In the case of conical structure,
our results are in accordance with those presented on reference \cite{Nano}.

*

\vspace{4.0cm}

\maketitle

{\large \bf Supplementary Information II}
\\
\\
{\large \bf Considerations regarding three dimensional rough LACFESs}
\\
\\
In this Supplementary Information section, we wish to demonstrate that the same relevant results obtained in the one-dimensional case (using orthodox cold field electron emission theory) are also obtained for a genuine rough two-dimensional LACFES. For this purpose, we use fractional Brownian motion (fBm) algorithms \cite{FBM} to generate random self-affine objects with specific Hurst exponents $H$. FBm algorithms can be generalized to higher dimensions using a multidimensional process such that for three dimensions, an irregular LACFES of height $h(x,y)$ results in interfaces that are statistically equivalent in all directions. We consider the self-affine sets that are constructed using the well-known midpoint displacement algorithm for $H$=0.1 \cite{Mid}.

\begin{figure}
\includegraphics [width=9.5cm] {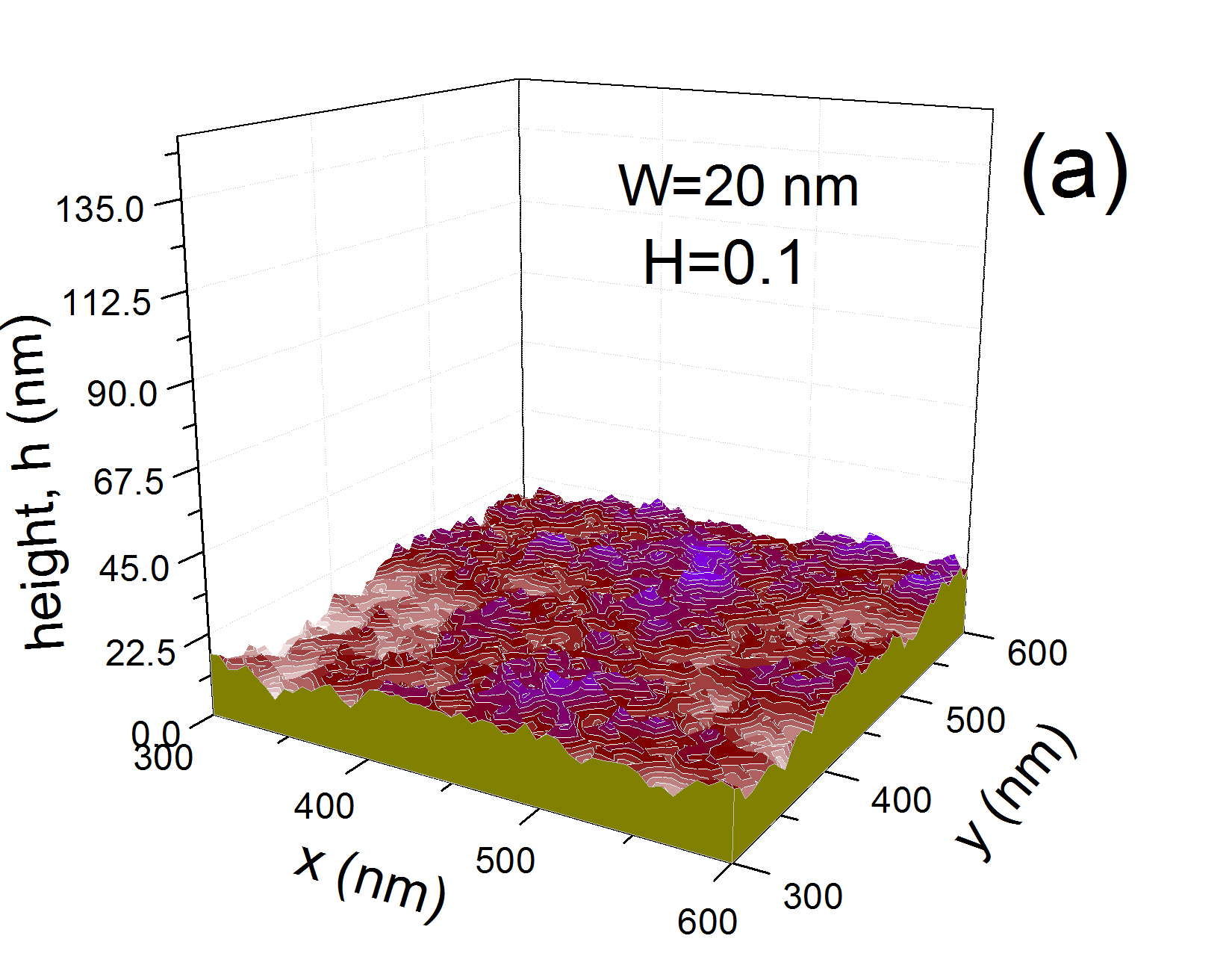}
\includegraphics [width=9.5cm] {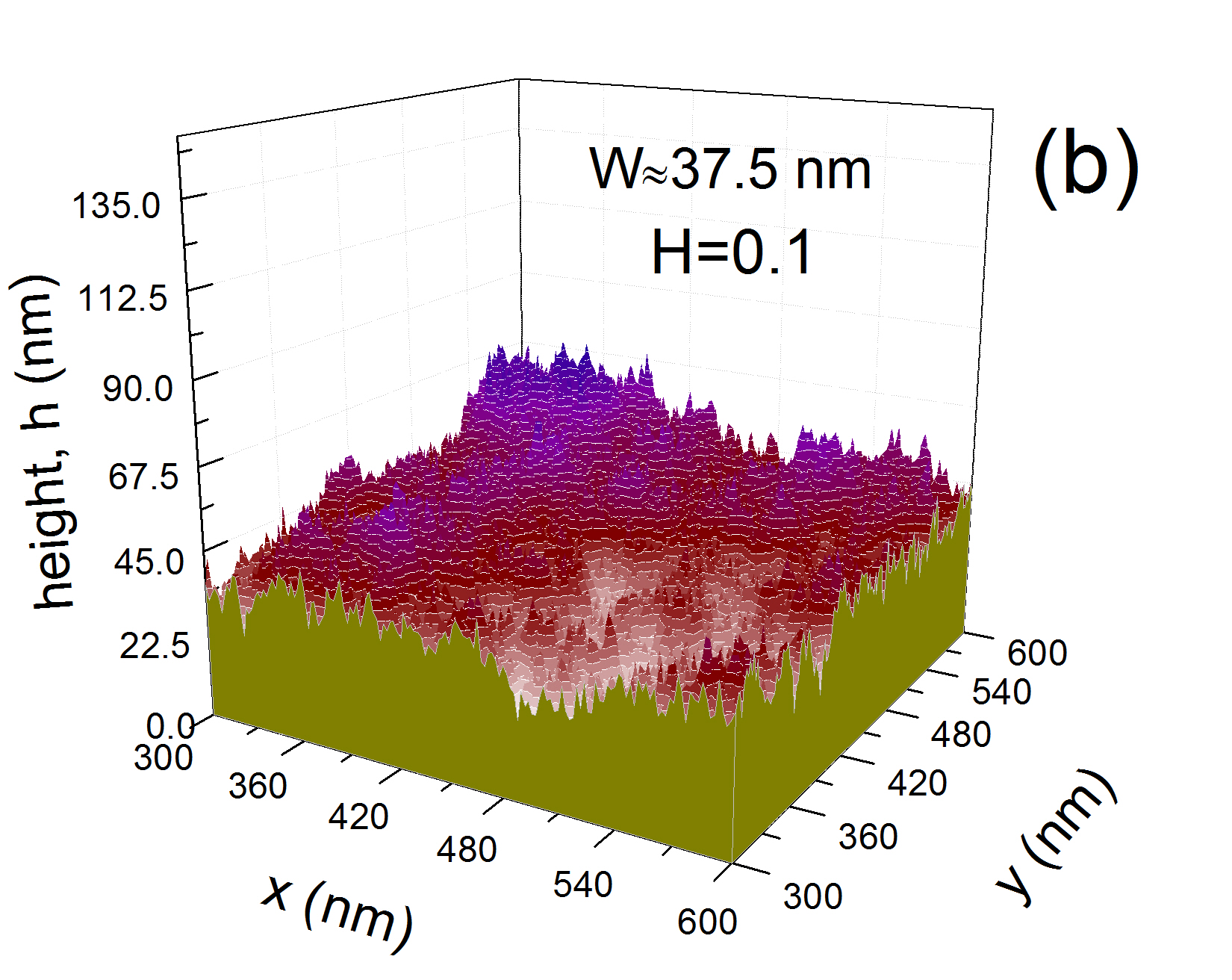}
\includegraphics [width=9.5cm] {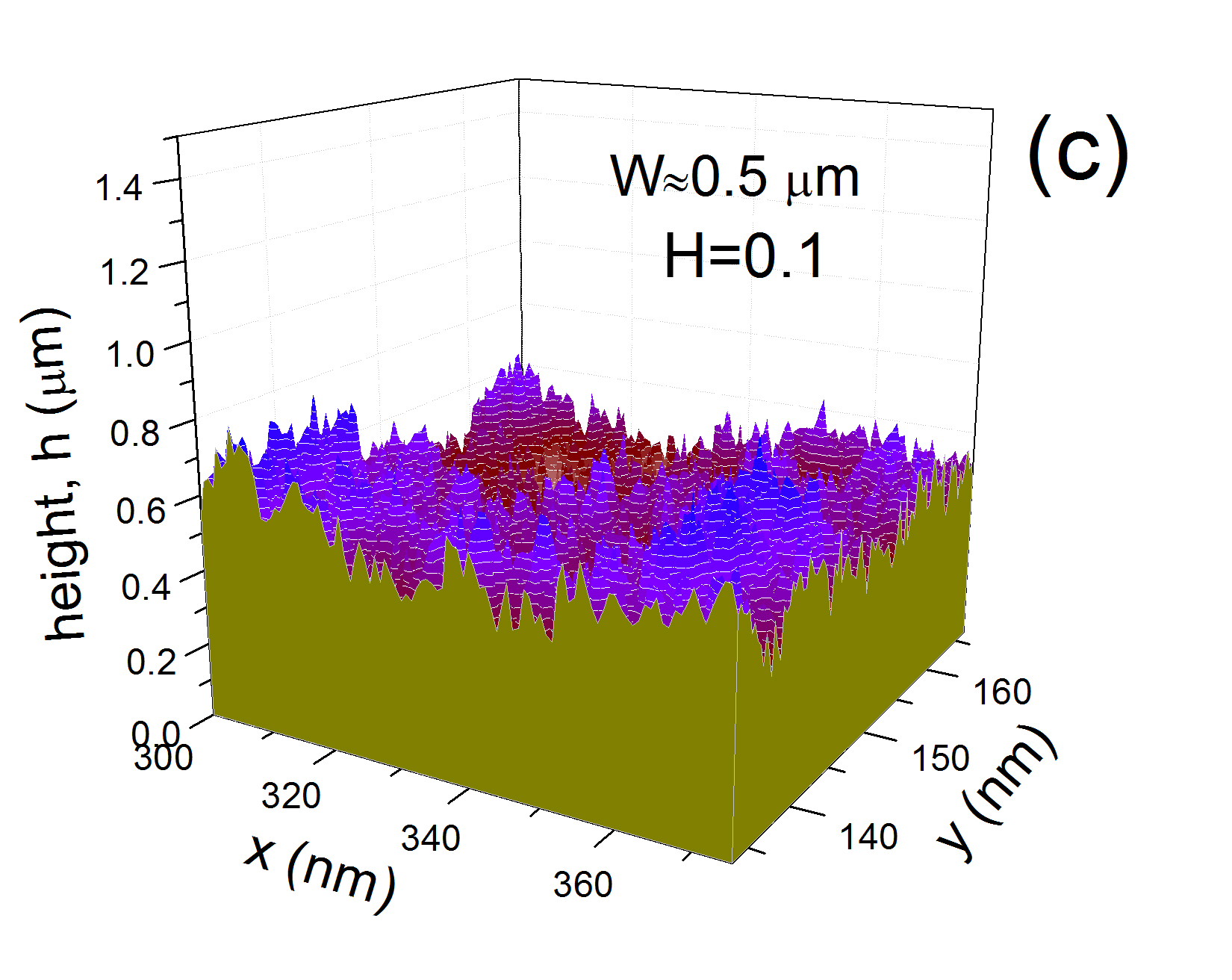}
\caption{Portion of rough LACFES generated using the fBm algorithm for a Hurst exponent of H=0.1 and global roughnesses: (a) $W \approx 20$ nm; (b) $W \approx 37.5$ nm and (c) $W \approx0.5\mu$m.}
\label{FBMH01}
\end{figure}

\begin{figure}
\includegraphics [width=15.5cm] {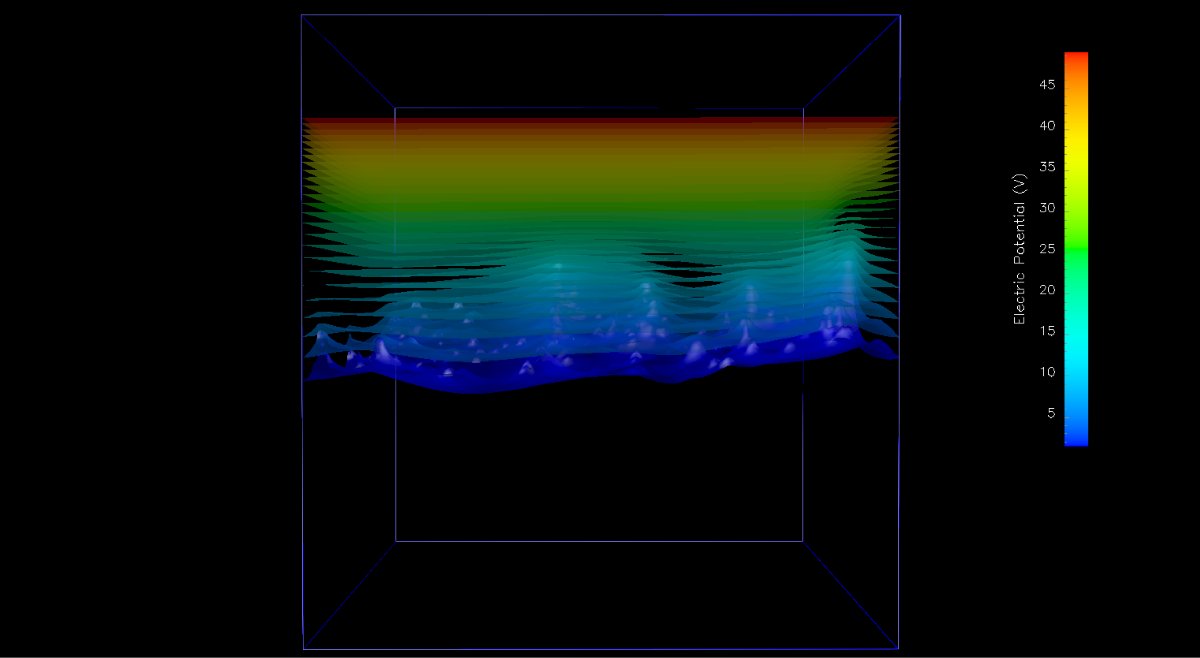}
\includegraphics [width=15.5cm] {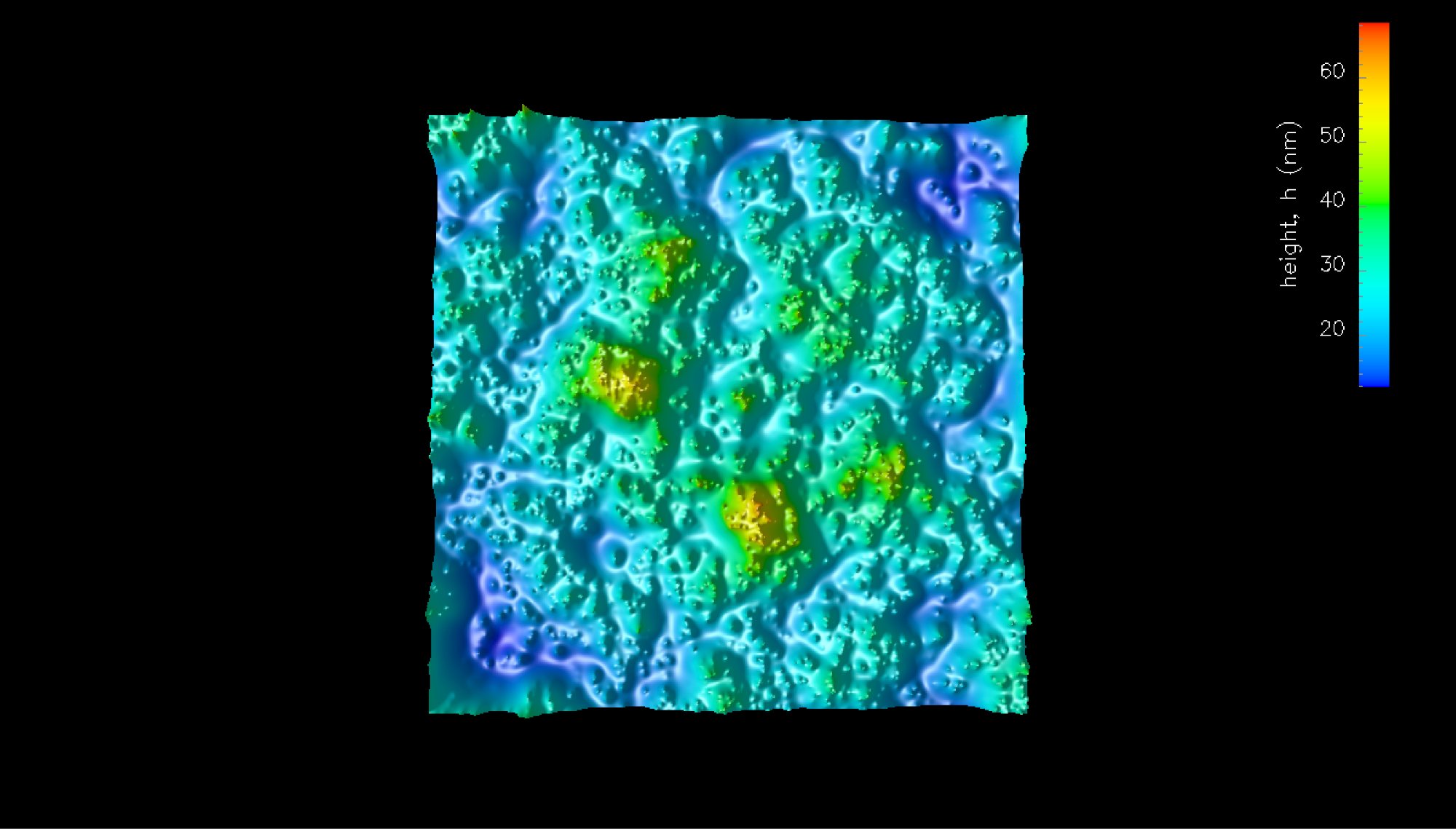}
\caption{
(top) Snapshot showing several equipotential surfaces calculated numerically from the solution of the Laplace equation, using the LACFES represented in Fig. \ref{FBMH01} (b) with appropriate Dirichlet conditions ($\Phi^{S} = 0$ V and $\Phi^{A} = 500$ V - see the text and Supplementary Information I for more details). The color bars indicate the electric potential values (in Volts). (bottom) The equipotential surface that is used as an approximation to the field emitter surface corresponding to the real fBm surface, defined by $\Phi^{E} = 1$ V. The color bars indicate the height of LACFES (in nanometers).}
\label{FBMH02}
\end{figure}

In Fig. \ref{FBMH01}, we illustrate the rough LACFESs produced using the described fBm algorithm for a Hurst exponent of $H$=0.1 and with global roughnesses $W \approx 20$ nm, $37.5$ nm and $0.5 \mu$m. In Fig. \ref{FBMH02}, several equipotential surfaces are shown for the case in which the electric potentials of the LACFES (with $H=0.1$ and $W \approx 37.5$ nm) and the far-away conducting anode are $\Phi^{S} = 0$ V and $\Phi^{A} = 500$ V, respectively, corresponding to Dirichlet conditions. We apply this approximation, following the same procedure used in the two-dimensional case, with the intent of rounding the sharpest projections. In this way, we replace the real fBm surface with an equipotential surface, $\Phi^{E} = 1$ V (very similar to the previous one), that represents an approximation of the field emitter surface. Again, following the same procedure using the WM function, this methodology ensures that on small scales (though larger than the atomic scale), the local roughness is negligible, whereas on large scales (though much less than the lateral size of the system, $L$), the local roughness scales with the same exponent $H$.

\begin{figure}
\includegraphics [width=15.5cm] {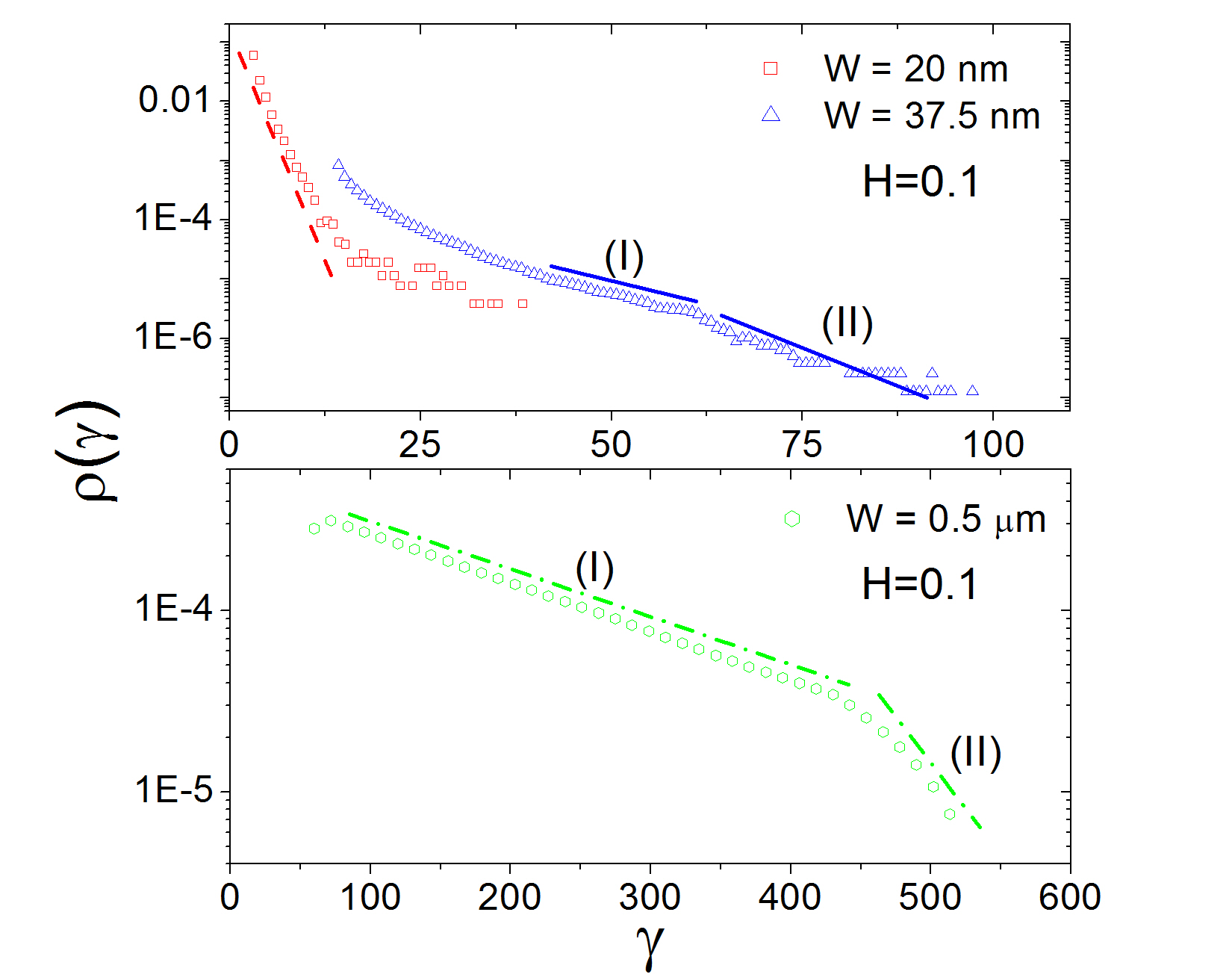}
\caption{Local FEF distributions, $\rho(\gamma)$, for LACFESs constructed using the fBm algorithm for $H = 0.1$ with global roughnesses of $W\approx20$ nm (red squares), $W\approx37.5$ nm (blue triangles) and $W\approx0.5\mu$m (green hexagons). Exponentially decaying behavior can be clearly observed in all cases. In particular, for $W\approx37.5$ nm and $W\approx0.5\mu$m, the same behavior previously observed, that of two exponentially decaying regimes, is apparent. The slope of the dashed (red) line is $(-0.247\pm0.008)$. The slopes of the solid (blue) lines are $(-0.0324\pm0.0001)$ (region (I)) and $(-0.046\pm0.001)$ (region (II)). The slopes of the dot-dashed (green) lines are $(-0.00273\pm0.00005)$ (region (I)) and $(0.0090\pm0.0001)$ (region (II)) (see Table III). The differences observed in the $\rho(\gamma)$ distributions are essentially related to the difference in global roughness between the LACFESs.}
\label{FBMH03}
\end{figure}

In Fig. \ref{FBMH03}, we present the local FEF distributions, $\rho(\gamma)$, for LACFESs shown in Fig. \ref{FBMH01}. Interestingly, exponential behaviors (two exponentially decaying regions as well as a characteristic FEF factor of the same order, for $W\approx37.5$ nm and $W\approx0.5 \mu$m, as that observed in thin-film emitters with irregular surfaces - see Ref.[30] of the manuscript) are observed in all cases, the same behavior as that discussed in the case of WM surfaces, when the one-dimensional field variation across the emitter surface is considered. Thus, the one-dimensional case is an interesting tool that captures the main results experimentally obtained for rough LACFESs, as noted previously.

Note that for $H=0.1$ and $W\approx37.5$ nm [$\approx 0.5\mu$m] (in which case the characteristic FEF has been found to be $\gamma_{C} \approx 97.24$ [$\approx 519.74$]), two exponentially decaying regimes are evident, labeled as (I) and (II), which correspond to $\rho^{I}(\gamma) \sim \exp(-\delta^I_{H} \gamma)$ and $\rho^{II}(\gamma) \sim \exp(-\delta^{II}_{H} \gamma)$, respectively, where $\delta^{II}_{H} > \delta^I_{H}$. For $W \approx 37.5$ [$\approx 0.5\mu$m], $\delta^{I}_{0.1} \times \log_{10} (e) = (0.0324 \pm 0.0001)$ [$\delta^{I}_{0.1} \times \log_{10} (e) = (0.00273\pm0.00005$)] is identified for interval (I), and $\delta^{II}_{0.1} \times \log_{10} (e) = (0.046 \pm 0.001)$ [$\delta^{II}_{0.1} \times \log_{10} (e) = (0.0090\pm0.0001)$] is found for interval (II). By contrast, for $W \approx 20$ nm, only one exponentially decaying regime is observed, with $\delta^{I}_{0.1} \times \log_{10} (e) = (0.247\pm0.008)$. This finding indicates that the $\rho(\gamma)$ distribution is also sensitive to changes in the global roughness of the LACFES. Differences between the theoretically obtained values of the characteristic FEF and those obtained experimentally can also be caused by differences in the global roughness. Table III summarizes the values of the parameters extracted from the $\rho(\gamma)$ distributions presented in Fig. \ref{FBMH03} and the $\gamma_C$ values for all values of $W$ explored.

\begin{figure}
\includegraphics [width=10.5cm] {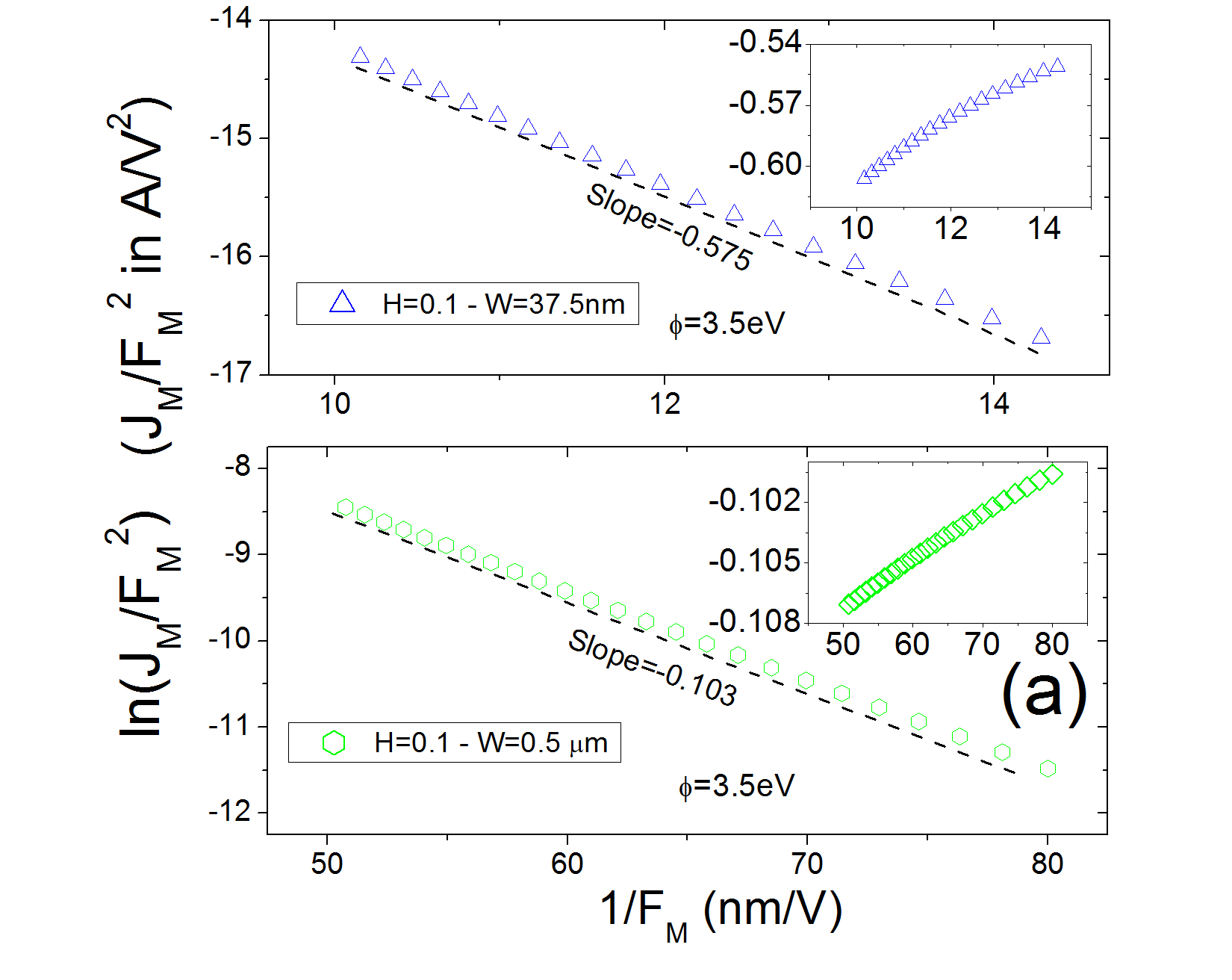}
\includegraphics [width=10.5cm] {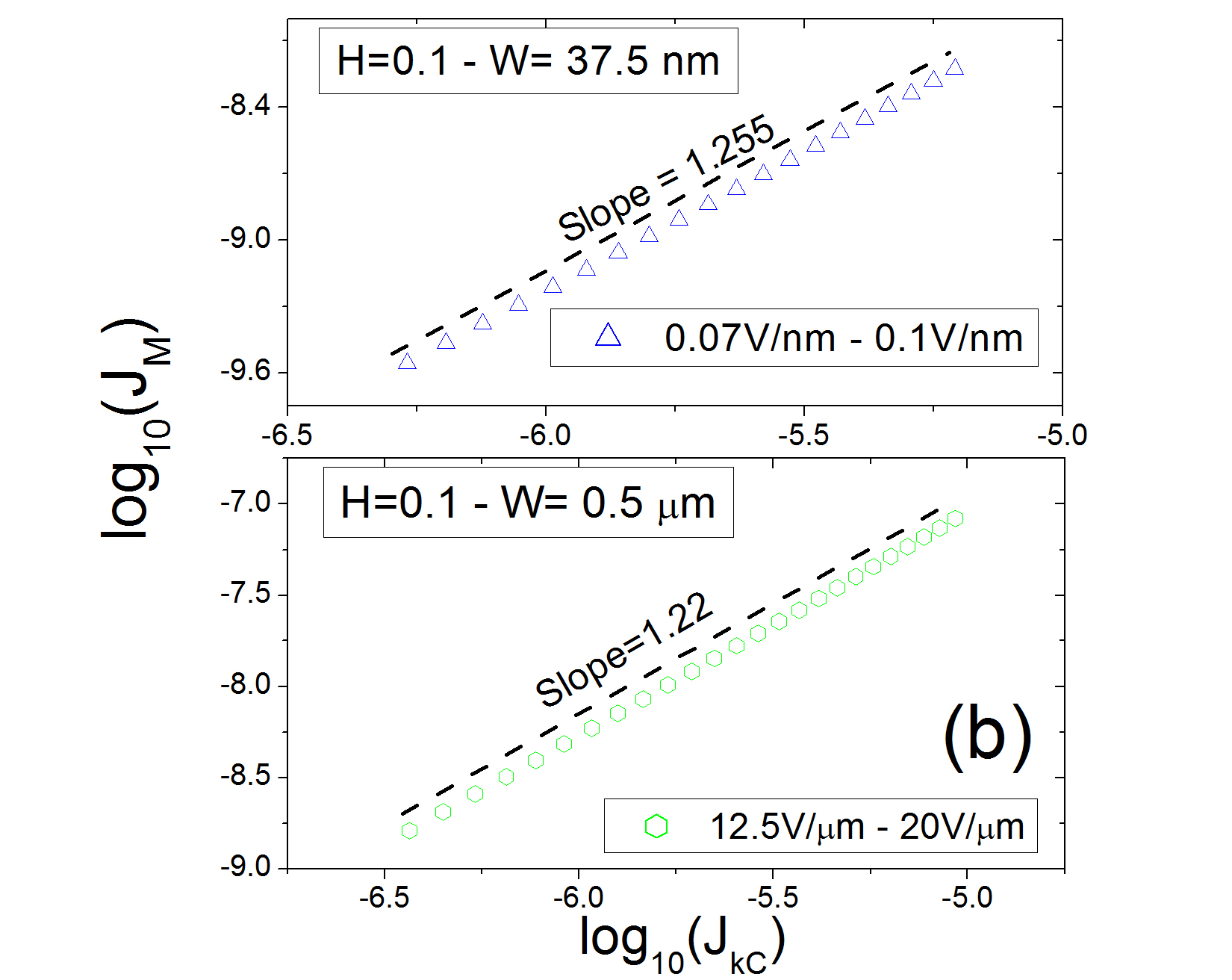}
\caption{(a) $J_{M}$-$F_{M}$-type FN plot for LACFESs with $H=0.1$ and $W\approx37.5$ nm (blue triangles) and $W\approx0.5\mu$m (green hexagons). The corresponding slopes (dashed lines) are shown (see Table IV). The work function of the LACFES is considered to be approximately constant and equal to $\phi = 3.5$ eV. The inset of this figure shows the corresponding derivatives of the $J_{M}$-$F_{M}$-type FN plots for LACFESs. (b) Macroscopic current density, $J_{M}$ as a function of the characteristic kernel current density, $J_{kC}$ with $H=0.1$ and $W\approx37.5$ nm (blue triangles) and $W\approx0.5\mu$m (green hexagons). For $W\approx37.5$ nm and $W\approx0.5\mu$m the macroscopic electric fields are in the range $0.07 V/nm \leq F_{M} \leq 0.1V/nm$ and $12.5 V/\mu m \leq F_{M} \leq 20 V/\mu$m, respectively. The slopes (values of $\beta_{W}$) are also indicated (see Table IV).}
\label{FBMH04}
\end{figure}

Finally, we show that the $\beta_W$ (note that the notation is now used with $W$ for global roughness) correction, which includes the effect of the geometry of the three-dimensional LACFES on the estimation of the characteristic FEF, must be considered to ensure more precise estimation of $\gamma_{C}$ (which was found to be $\gamma_{C} \approx 97.24$ [$\approx 519.74$] from the $\rho(\gamma)$ distribution depicted in Fig. \ref{FBMH03} for $W \approx 37.5$ nm [$\approx 0.5\mu$m]). We restrict this discussion to $H=0.1$ and for roughnesses $W\approx 37.5$ nm and $W\approx 0.5 \mu$m, i.e., to the LACFESs represented in Figs. \ref{FBMH01} (b) and (c), respectively. Fig. \ref{FBMH04}(a), shows the behavior of the corresponding $J_{M}$-$F_{M}$-type FN plots. The FN plots seems, again, to exhibit approximately linear behavior for the considered range of macroscopic electric field $F_M$, although, in reality, the derivative is not constant, as observed in the corresponding inset.

Fig. \ref{FBMH04}(b) shows the behavior of $\log_{10}(J_{M})$ as a function of $\log_{10}(J_{kC})$, considering a macroscopic electric field in the range $0.07 V/nm \leq F_{M} \leq 0.1V/nm$ and $12.5 V/\mu m \leq F_{M} \leq 20 V/\mu$m for $W\approx37.5$ nm and $W\approx0.5\mu$m, respectively. These results suggest, again, a scaling relation between $J_{M}$ and $J_{kC}$ with $\beta_W = (1.255 \pm 0.001)$ for $W\approx37.5$ nm [and $\beta_W = (1.222 \pm 0.002)$ for $W\approx0.5\mu$m]. From these results, for $W\approx37.5$ nm [$\approx0.5\mu$m] we calculate an elementary slope characterization parameter of $\gamma_{C}^{T}\approx 77.7$ [$\approx434$] (error of 21\% [16.4 \%] with respect to the $\gamma_C$ value found from the $\rho(\gamma)$ distribution). If we include the corrections $\sigma$ (which is a generalized slope correction factor related to the SN barrier) and $\beta_W$ (which includes the effect of the global roughness of the LACFES on the estimation of the characteristic FEF), then the corrected slope characterization parameter is found to be $\gamma^{\beta_W,\sigma}_{C} \approx 93$ [$\approx 505.8$], yielding an error of 4\% [2.6\%] with respect to the $\gamma_C$ value determined from the $\rho(\gamma)$ distribution (see Table IV for values of the slopes of the $J_{M}$-$F_{M}$-type FN plots shown in Fig. \ref{FBMH04} (a), the characteristic FEF values $\gamma_{C}^{T}$, the $\beta_{W}$ values extracted from the results presented in Fig.\ref{FBMH04} (b) and $\gamma^{\beta_W,\sigma}_{C}$).

\begin{table}
   \centering
 \renewcommand{\arraystretch}{1.5}


 \begin{tabular}{|c|c|c|c|}

  \hline
 $W$ & $\delta_{H}^{I}$ $\times \log_{10} (e)$ & $\delta_{H}^{II}$ $\times \log_{10} (e)$ & $\gamma_{C}$ \\

 \hline

 $\approx 20$ nm & $0.247\pm0.008$ & $-$ &  $\approx 38.35$   \\ \hline
 $\approx 37.5$ nm & $0.0324\pm0.0001$ & $0.046\pm0.001$  & $\approx 97.24$  \\ \hline
 $\approx 0.5 \mu$m & $0.00273\pm0.00005$ & $0.0090\pm0.0001$  & $\approx 519.74$ \\ \hline

 \end{tabular}

 \caption{Results of the extraction of parameters from the $\rho(\gamma)$ distribution shown on Fig. \ref{FBMH03} for $H=0.1$ and global roughnesses $W \approx 20$ nm, $37.5$ nm and 0.5 $\mu$m. For $W \approx 20$ nm a dominant exponential decay can be observed in the $\rho(\gamma)$ distribution, characterized by $\delta_{H}^{I}$.}
 \label{tab}
 \end{table}

In conclusion, the scale that results in the exponent $\beta_W$ must be considered (including in the case of a three-dimensional setup and in addition to the generalized slope correction from the SN barrier) in the calculation of the slope characterization parameter that is used by experimentalists to extract a more precise characteristic FEF, such as that found using the one-dimensional WM function.

\begin{table}
   \centering
 \renewcommand{\arraystretch}{1.5}


 \begin{tabular}{|c|c|c|c|c|}

  \hline
 $W$ & $S_{F_M}(W) (V/nm)$ & $\gamma_{C}^{T}$ & $\beta_W$ & $\gamma^{\beta_W,\sigma}_{C}$  \\

 \hline

$\approx37.5$ nm & $-0.5759 \pm 0.0009$ & $77.77 \pm 0.01$ & $1.255 \pm 0.001$ & $93.2 \pm 0.3$   \\ \hline
$\approx 0.5 \mu$m & $-0.1031\pm 0.002$ & $434.2 \pm 0.1$ & $1.224 \pm 0.001$ & $505.88 \pm 0.01$ \\ \hline
\end{tabular}

\caption{Results for $H=0.1$ and $W \approx 20$ nm, $37.5$ nm and $0.5 \mu$m. The slopes of the $J_{M}$-$F_{M}$-type FN plots shown on Fig.\ref{FBMH04} (a), characteristic FEF $\gamma_{C}^{T}$, the $\beta_{W}$ values extracted from the results presented on Fig. \ref{FBMH04}(b) and $\gamma^{\beta_W,\sigma}_{C}$. $\gamma_{C}^{T}$ and $\gamma^{\beta_W,\sigma}_{C}$ were calculated using Eqs. (21) and (23) while considering the elementary FN equation and relevant corrections, including the SN barrier and the morphology of the LACFES, respectively.}
\label{tab}
\end{table}

%

%

\end{document}